\documentclass[zpreprint,zbstdefault]{./LaTeX/zeus/zeus_paper}
\usepackage[english]{babel}
\chardef\usc=95
\chardef\til=126
\catcode`\@=11 
\DeclareRobustCommand\xdotspace{\futurelet\@let@token\@xdotspace}
\def\@xdotspace{%
  \ifx\@let@token.\else
  \ifx\@let@token\bgroup.\else
  \ifx\@let@token\egroup.\else
  \ifx\@let@token\/.\else
  \ifx\@let@token\ .\else
  \ifx\@let@token~.\else
  \ifx\@let@token!.\else
  \ifx\@let@token,.\else
  \ifx\@let@token:.\else
  \ifx\@let@token;.\else
  \ifx\@let@token?.\else
  \ifx\@let@token/.\else
  \ifx\@let@token'.\else
  \ifx\@let@token).\else
  \ifx\@let@token-.\else
  \ifx\@let@token\@xobeysp.\else
  \ifx\@let@token\space.\else
  \ifx\@let@token\@sptoken.\else
   .\space
   \fi\fi\fi\fi\fi\fi\fi\fi\fi\fi\fi\fi\fi\fi\fi\fi\fi\fi}
\catcode`\@=12 

\newcommand{\stru}[2]{%
   \relax\ifmmode\hbox{\vrule height#1 depth#2 width0pt}%
   \else\vrule height#1 depth#2 width0pt\fi}

\newcommand{\Ronum}[1]{\uppercase\expandafter{\romannumeral#1}}
\newcommand{\ronum}[1]{\expandafter{\romannumeral#1}}
\DeclareRobustCommand{\LaTeXZ}{%
  \LaTeX\kern-.05em4\kern-.1em
  {\raisebox{-0.2ex}{$\scriptstyle\text{ZEUS}$}}\xspace}

\DeclareMathAlphabet{\mathbf}{OT1}{cmr}{bx}{sl}
\newcommand{\eVdist}{\kern-0.06667em}

\newcommand{\slashfrac}[2]{%
  \raisebox{0.5ex}{\ensuremath #1}\kern-0.12em/\kern-0.08em
  \raisebox{-.8ex}{\ensuremath #2}}

\newcommand{\sqr}[3]{%
    {\vcenter{\hrule height.#3ex\hbox{\vrule width.#2ex height#1ex
     \kern#1ex\vrule width.#3ex}\hrule height.#2ex}}}

\catcode`\@=11 
\newcommand{\parenbar}{\mathpalette\p@renb@r}
\def\p@renb@r#1#2{\vbox{%
  \ifx#1\scriptscriptstyle \dimen@.7em\dimen@ii.2em\else
  \ifx#1\scriptstyle \dimen@.8em\dimen@ii.25em\else
  \dimen@1em\dimen@ii.4em\fi\fi \offinterlineskip
  \ialign{\hfill##\hfill\cr
    \vbox{\hrule width\dimen@ii}\cr
    \noalign{\vskip-.3ex}%
    \hbox to\dimen@{$\mathchar300\hfil\mathchar301$}\cr
    \noalign{\vskip-.3ex}%
    $#1#2$\cr}}}
\catcode`\@=12 

\newcommand{\IP}{{\rm I$\kern-0.01667em$P}\xspace}

\mathchardef\qsm=63
\mathchardef\pls=43
\mathchardef\mns=512
\mathchardef\plm=518
\mathchardef\eql=61
\mathchardef\smallleft=300
\mathchardef\smallright=301
\mathchardef\les=316
\mathchardef\gre=318
\mathchardef\leq=532
\mathchardef\grq=533

\catcode`\@=11 
\newcounter{pict@width}
\newcounter{pict@height}
\newlength{\pict@scale}
\newcommand{\psfigadd}[4]{%
\setcounter{pict@width}{1*\ratio{#2+\pict@scale/2}{\pict@scale}}
\setcounter{pict@height}{1*\ratio{#3+\pict@scale/2}{\pict@scale}}
\setlength{\unitlength}{\pict@scale}
\hbox to #2{\hspace{-\fill}\begin{picture}(\thepict@width,\thepict@height)
\put(0,0){\psfig{figure=#1,width=#2,height=#3,clip=}}
\SetScale{0.283466457}
\SetWidth{1.763889}
{#4}
\end{picture}}
}
\newcounter{pict@widthfst}
\newcounter{pict@widthscd}
\newcounter{pict@widthtot}
\newcommand{\psfigaddtwo}[7]{%
\setcounter{pict@widthfst}{1*\ratio{#2+\pict@scale/2}{\pict@scale}}
\setcounter{pict@widthscd}{1*\ratio{#2+#4+\pict@scale/2}{\pict@scale}}
\setcounter{pict@widthtot}{1*\ratio{#2+#4+#6+\pict@scale/2}{\pict@scale}}
\setcounter{pict@height}{1*\ratio{#3+\pict@scale/2}{\pict@scale}}
\setlength{\unitlength}{\pict@scale}
\hbox{\hspace{-\fill}\begin{picture}(\thepict@widthtot,\thepict@height)
\put(0,0){\psfig{figure=#1,width=#2,height=#3,clip=}}
\put(\thepict@widthscd,0){\psfig{figure=#5,width=#6,height=#3,clip=}}
\SetScale{0.283466457}
\SetWidth{1.763889}
{#7}
\end{picture}}
}
\newcommand{\psfigror}[4]{%
\setcounter{pict@width}{1*\ratio{#2+\pict@scale/2}{\pict@scale}}
\setcounter{pict@height}{1*\ratio{#3+\pict@scale/2}{\pict@scale}}
\setlength{\unitlength}{\pict@scale}
\hbox{\begin{picture}(\thepict@width,\thepict@height)
\put(0,\thepict@height){\psfig{figure=#1,width=#3,height=#2,clip=,angle=270}}
\SetScale{0.283466457}
\SetWidth{1.763889}
{#4}
\end{picture}}
}
\newcommand{\psfigrol}[4]{%
\setcounter{pict@width}{1*\ratio{#2+\pict@scale/2}{\pict@scale}}
\setcounter{pict@height}{1*\ratio{#3+\pict@scale/2}{\pict@scale}}
\setlength{\unitlength}{\pict@scale}
\hbox{\begin{picture}(\thepict@width,\thepict@height)
\put(0,0){\psfig{figure=#1,width=#3,height=#2,clip=,angle=90}}
\SetScale{0.283466457}
\SetWidth{1.763889}
{#4}
\end{picture}}
}
\catcode`\@=12 
\newlength\listtextwidth

\catcode`\@=11 
\newlength{\@tabfninsert}
\newlength{\@tabfnwidth}
\newcommand{\tabfootnote}[2]{%
  \setlength{\@tabfninsert}{0.8em}
  \setlength{\@tabfnwidth}{\textwidth}
  \addtolength{\@tabfnwidth}{-\@tabfninsert}
  \addtolength{\@tabfnwidth}{-0.4em}
  \noindent\makebox[\@tabfninsert][r]{\footnotesize$^{#1}$\hfil}\hfill%
  \parbox[t]{\@tabfnwidth}{\footnotesize #2\hfill}}
\catcode`\@=12 

\begin{document}
\newcommand{\lw}[1]{\smash{\lower1.3ex\hbox{#1}}}
\prepnum{DESY--02--105}
\newcommand{\asmz}{\alpha_s(M_Z)}

\title{
A ZEUS next-to-leading-order QCD analysis of data on deep inelastic scattering
}                                                       
                    
\author{ZEUS Collaboration}

\date{}
\abstract{
   Next-to-leading-order QCD analyses of the ZEUS data on 
   deep inelastic scattering 
   together with fixed-target data have been performed, from which 
   the gluon and quark densities of the proton and the value of the strong
   coupling constant, 
   $\asmz$, were extracted. The study includes a full
   treatment of the experimental systematic uncertainties 
   including point-to-point correlations.
   The resulting uncertainties in the parton density functions 
   are presented. A combined fit 
   for $\asmz$ and the gluon and quark densities yields a value for $\asmz$
   in agreement with the world average.
    The parton density functions derived from ZEUS data alone 
    indicate the importance of HERA data 
    in determining the sea quark and gluon distributions at
    low $x$.
    The limits of applicability of the theoretical formalism have been 
    explored by
    comparing the fit predictions to ZEUS data at very low $Q^2$.    
}

\makezeustitle

\newcommand{\address}{ }                                                                           
\pagenumbering{Roman}                                                                              
                                                   %
  S.~Chekanov,                                                                                     
  D.~Krakauer,                                                                                     
  S.~Magill,                                                                                       
  B.~Musgrave,                                                                                     
  J.~Repond, and                                                                                      
  R.~Yoshida\\                                                                                     
 {\it Argonne National Laboratory, Argonne, Illinois 60439-4815}                            
\par \filbreak                                                                                     
  M.C.K.~Mattingly \\                                                                              
 {\it Andrews University, Berrien Springs, Michigan 49104-0380}                                    
\par \filbreak                                                                                     
  P.~Antonioli,                                                                                    
  G.~Bari,                                                                                         
  M.~Basile,                                                                                       
  L.~Bellagamba,                                                                                   
  D.~Boscherini,                                                                                   
  A.~Bruni,                                                                                        
  G.~Bruni,                                                                                        
  G.~Cara~Romeo,                                                                                   
  L.~Cifarelli,                                                                                    
  F.~Cindolo,                                                                                      
  A.~Contin,                                                                                       
  M.~Corradi,                                                                                      
  S.~De~Pasquale,                                                                                  
  P.~Giusti,                                                                                       
  G.~Iacobucci,                                                                                    
  A.~Margotti,                                                                                     
  R.~Nania,                                                                                        
  F.~Palmonari,                                                                                    
  A.~Pesci,                                                                                        
  G.~Sartorelli, and                                                                                   
  A.~Zichichi  \\                                                                                  
  {\it University and INFN Bologna, Bologna, Italy}
% ...
\newpage                                                                                           
% ...
                                                           %

                                                           %

\pagenumbering{arabic} 
\pagestyle{plain}

\newcommand{\unit}[1] {\mbox{\hspace{0.3em}\rm #1}}
\newcommand{\msbar}{\mbox{$\overline{\rm{MS}}$}\ }
\newcommand{\leqsim}{\,\raisebox{-0.6ex}{$\buildrel < \over \sim$}\,}
\newcommand{\geqsim}{\,\raisebox{-0.6ex}{$\buildrel > \over \sim$}\,}
\section{Introduction}
\label{sec:intro}

Studies of inclusive differential cross sections and structure functions, 
as measured in deep inelastic scattering (DIS) of leptons from hadron
targets, played a crucial role in
establishing the theory of perturbative quantum chromodynamics (pQCD).
The next-to-leading-order (NLO) DGLAP evolution 
equations~\cite{sovjnp:15:438,sovjnp:20:94,np:b126:298,jetp:46:641}
form the basis for a successful description 
of the data over a broad kinematic range. Thus parton distribution 
functions (PDFs) and the value of the strong coupling constant, $\asmz$, can be
determined within this formalism. 
The availability of HERA data has greatly increased the 
kinematic range over which such studies can be made. 

The MRST~\cite{epj:c23:73} and 
CTEQ~\cite{hep-ph-0201195} groups have used the most recent HERA 
data~\cite{epj:c21:443,epj:c21:33} in global fits to determine PDFs 
and $\asmz$.
In recent years, estimating the uncertainties on PDFs from experimental 
sources, as well
as from model assumptions, has become an 
issue~\cite{pr:d58:094023,epj:c14:285,hep-ex-0005042,hep-ph-0006148,pr:d65:014012,pr:d65:014013,hep-ph-0104052,jp:g28:779}.
The CTEQ group 
has made a detailed study of the uncertainties  
on the PDFs due to experimental sources, whereas MRST provide
four sets of PDFs from fits done with different theoretical assumptions.
The best fits of these groups differ somewhat, reflecting differences of
approach. The H1 collaboration has also considered the uncertainties on 
the gluon distribution and $\asmz$ resulting from a fit to H1 and BCDMS 
data~\cite{epj:c21:33}. 

In this paper, the ZEUS data from 1996 and 1997~\cite{epj:c21:443} 
have been used, together with fixed-target data, to extract
gluon and sea densities with much improved precision compared to
earlier work that used the ZEUS 1994/95 data~\cite{epj:c7:609,epj:c14:285}. 
The fixed-target data are important for  
a precise determination of the valence distributions. 
 All parton distributions have been 
extracted taking into account the point-to-point correlated systematic
uncertainties of the input data. 

The value of $\asmz$ was
set to the world-average value, $\asmz = 0.118$~\cite{epj:c15:1},
for the determination of parton distributions 
in the standard fit (called ZEUS-S).
The increased precision of the data also allows 
a determination of the value of $\asmz$. The
correlations between the shape of the parton distribution functions and the 
value of $\asmz$ have been fully 
taken into account by making a simultaneous fit
to determine the parton distribution parameters and $\asmz$. This fit is called
ZEUS-$\alpha_s$.

One of the main topics of this paper is an evaluation of the experimental 
uncertainties
on the extracted parton distribution functions and on the value of $\asmz$.
The treatment of point-to-point correlated systematic uncertainties
 reflects knowledge that
such uncertainties are not always Gaussian distributed.
Model and theoretical uncertainties have also been estimated. 

The role of ZEUS data has been explored by making a fit using ZEUS data alone.
The ZEUS charged current $e^+p$ data from 
1994-97~\cite{epj:c12:411}, and the charged and neutral current $e^-p$
data from the 1998/99 runs~\cite{pl:b539:197,desy:02:113}
were
used, together with the 1996/97 $e^+p$ neutral current data, to make 
an extraction of the PDFs independently of other
experiments. This fit is called ZEUS-O.

The extent to which the NLO DGLAP
formalism  continues to provide a successful description of the data over an 
increased kinematic range was investigated by comparing the ZEUS-S
fit to the ZEUS
high-precision data at very low $Q^2$~\cite{pl:b487:53}.
The combination of the improved fit analysis and the 
increased precision of these data, compared to those used~\cite{pl:b407:432} 
in the previous study~\cite{epj:c7:609}, 
allows a low-$Q^2$ limit to be put on the applicability of the 
NLO DGLAP description of DIS data.

The paper is organised as follows. In Section~\ref{sec:theory}, 
some theoretical background is given. In Section~\ref{sec:qcd}, the
NLO QCD fits to the ZEUS data and fixed-target DIS data are described, 
paying particular attention to the treatment of experimental uncertainties.
In Section~\ref{sec:standard}, the standard ZEUS-S fit is compared to data
and the extracted parton distribution functions including their 
experimental uncertainties are presented. The analysis is extended to evaluate
 $\asmz$ in the ZEUS-$\alpha_s$ fit and uncertainties from experimental and
theoretical sources are discussed. 
In Section~\ref{sec:special}, parton densities from the ZEUS-O fit are
presented and, in Section~\ref{sec:loq2qcd}, the 
limitations of the NLO DGLAP formalism are considered. 
Section~\ref{sec:concl} contains a summary and conclusions. In the Appendix,
various ways of treating systematic uncertainties are discussed and compared. 

\section{Theoretical Perspective}
\label{sec:theory}

The differential cross section for neutral current (NC) $e^+p$
deep inelastic scattering (DIS) is given in terms of the
structure functions $F_2$, $F_L$ and $xF_3$ by
\[
\frac {d^2\sigma (e^{+}p) } {dxdQ^2} =  \frac {2\pi\alpha^2} {Q^4 x}
\left[Y_+\,F_2^{ep}(x,Q^2) - y^2 \,F_L^{ep}(x,Q^2)
- Y_-\, xF_3^{ep}(x,Q^2) \right].
\]
The kinematic variables are 
Bjorken's $x=Q^2/(2p\cdot q)$ and the negative 
invariant-mass squared of the exchanged virtual
boson, $Q^2=-q^2$, 
where $p$ is the four-vector of the target proton
and $q=k-k'$ is the difference of the four-vectors of the incoming and 
outgoing leptons. The variable $y$ is defined by 
$y=(p\cdot q)/(p\cdot k)$ and $\displaystyle Y_\pm=1\pm(1-y)^2$. 
It is also useful to define $W^2$,
the virtual boson-proton squared centre-of-mass energy, given by
\[W^2 = (p+q)^2 = Q^2\frac{(1-x)}{x} + m_p^2,
\] 
where $m_p$ is the proton mass. The reduced 
cross section is defined as 
\[ \tilde{\sigma} = \frac {d^2\sigma (e^{+}p) }{dxdQ^2} \frac {Q^4 x}{2\pi\alpha^2 Y_+},
\] 
so that it is equal to $F_2$ 
when $F_L$ and $xF_3$ are negligible. For $Q^2$ values much below the 
$Z^0$-mass squared, the parity-violating structure function, $xF_3$, is 
negligible, since the cross sections are dominated 
by virtual photon exchange. Then, provided that $W^2$ is large enough that
target-mass and higher-twist contributions may be neglected, 
the structure function $F_2$ can be 
simply interpreted in LO QCD as the
sum of the quark distribution functions weighted with the quark-charges 
squared. To the same approximation, $F_L$, the longitudinal structure 
function, is zero.
At NLO, these structure functions are related to the
parton distributions in the proton through convolution with the QCD coefficient
functions. Since the ZEUS data extend to high $Q^2$,
the coefficient functions also include $Z^0$ 
exchange~\cite{roberts:1990:struc}.  
Measurement of the structure functions as a
function of $x$ and $Q^2$ yields information on the shape 
 of the parton distributions and, through their $Q^2$ dependence, on the value
of the strong coupling constant, $\asmz$.  

Before HERA data became available, leading-twist perturbative expansions of 
QCD, as formulated in the DGLAP evolution equations, were found to 
describe fixed-target data adequately down to $Q^2 \sim 4\unit{GeV}^2$ and 
$x \sim 10^{-2}$.  
The QCD evolution was typically started from $Q^2_0 \sim 4\unit{GeV}^2$ 
or higher.  
Convenient functional forms of the parton distribution functions (PDFs) were 
input, at $Q^2_0$, and fitted to the data.  
At small $x$, these were $xf(x) \simeq Ax^\delta$. The fits 
gave $\delta \sim 0.5$ for valence distributions and $\delta \sim 0$ (flat) 
for the sea and the gluon distributions.  

It was shown~\cite{pr:d10:1649} as early as 1974 that, according to QCD, this  
behaviour cannot persist to infinitely small values of $x$. 
At some point, a much steeper rise of the gluon distribution is expected, 
leading to a steeply rising behaviour of $F_2$.  However, 
it was unclear at what $x$ values such behaviour should begin.
Hence, prior to HERA operation, most predictions specified
a gentle rise at small $x$, as expected from soft Regge-like behaviour.  
The dramatic rise in $F_2$ observed in the early HERA  
data~\cite{pl:b316:412,np:b407:515}, at $x \sim 10^{-3}$,  
$Q^2 \sim 15\unit{GeV}^2$ was therefore a  surprise. Furthermore, later HERA 
data~\cite{zfp:c72:399,np:b470:3} showed that this rise persisted  
down to surprisingly low $Q^2$, of order 
$Q^2 \sim 1.5\unit{GeV}^2$, where $x \sim 5 \times 10^{-5}$.  
Applications of pQCD using the NLO DGLAP
formalism to data in this kinematic region
have been reasonably successful, although 
there are several issues that could limit the applicability of 
this formalism.

One question is whether only a few terms in the perturbative expansion are 
adequate, given the large values of $\alpha_s$ at low $Q^2$.
 Recent NLO DGLAP fits including HERA data have used starting 
values as low as $Q^2_0 \sim 1\unit{GeV}^2$.  
These fits have sea input distributions
 with $\delta \sim -0.2$. However, such fits require the gluon input 
distributions to be valence-like~\cite{pl:b387:419,epj:c4:463}, or  even 
negative~\cite{epj:c23:73}, at small $x$. This
calls into question the applicability of the DGLAP formalism at these low
values of $Q^2$.

Furthermore, at the low $x$ values accessed at HERA, large $\ln(1/x)$ terms, 
which are not included in the DGLAP formalism, could be important.  If so, the 
treatment may need to be amended by consideration of BFKL 
dynamics~\cite{sovjnp:23:338,jetp:45:199, sovjnp:28:822,sovjnp:63:904,pl:b429:127,pl:b430:349}.  

Finally, the high gluon density observed at higher $Q^2$ could lead to gluons 
screening each other from the virtual-boson probe, requiring 
non-linear terms in the evolution equations. These act oppositely to the 
linear terms, such that gluon evolution is slowed down and may even 
saturate~\cite{hep-ph-0111244}.  
 
It is unclear where any of these effects
become important.  Presently, the range of applicability of the 
NLO QCD expansion is a matter to be resolved by experiment.  
To draw firm conclusions requires precision data and a 
careful analysis of the uncertainties on the predicted shapes of the parton 
distributions.  In the present paper, the high-precision data from the ZEUS 
experiment and all fixed-target experiments for which full information on 
correlated systematic uncertainties is available have been used to
extract the PDFs and $\asmz$ and to 
investigate the range of applicability of the NLO DGLAP formalism.

\section{ Description of NLO QCD fits}
\label{sec:qcd}
This section gives the specifications of the ZEUS-S and ZEUS-$\alpha_s$ 
 global NLO QCD fits to the new ZEUS cross-section 
data~\cite{epj:c21:443} and fixed-target DIS data.

The fixed-target data were included to constrain the fits at high $x$ 
and provide information on the
valence distributions and the flavour composition of the sea.
All high-precision fixed-target data sets for which full information on the
correlated systematic uncertainties is available have been used:
\begin{itemize}
\item
$F_2$ data on $\mu-p$ scattering from BCDMS~\cite{pl:b223:485}, 
NMC~\cite{np:b483:3}, and E665~\cite{pr:d54:3006};
\item
deuterium-target data 
from NMC~\cite{np:b483:3} and E665~\cite{pr:d54:3006}. These were
included in order to have $\bar u$, $\bar d$ flavour separation; 
\item
NMC data on  the ratio $F_2^D/F_2^p$~\cite{np:b487:3}. These determine
the ratio of the  $d$ to $u$ valence shapes;
\item
the CCFR~\cite{prl:79:1213} $xF_3$ 
data, from (anti-)neutrino interactions on an iron target. These 
give the strongest
constraint on high-$x$ valence PDFs.  They are used only in the $x$ range 
$ 0.1 ~\leq ~x ~\leq ~0.65$ in order to minimize
dependence on the heavy-target corrections. 
The latter were performed according to 
the prescription of MRST~\cite{epj:c4:463}.
These $xF_3$ data are unaffected by the recent 
re-analysis of CCFR $F_2$ data~\cite{hep-ex-0011095,hep-ex-0010001}.
\end{itemize}

The deuterium data were corrected
to represent $(n+p)/2$ by the prescription of Gomez et al.~\cite{pr:d49:4348}.
The fit results were found to be insensitive to
the specific prescriptions used for heavy-target and deuterium corrections.

The fits were performed at leading twist. The following cuts were made on
the ZEUS and the fixed-target data: 
\begin{itemize}
  \item $Q^2 > 2.5\unit{GeV}^2$ was required 
 to remain in the kinematic region where
perturbative QCD is expected to be applicable;
  \item $W^2  > 20\unit{GeV}^2$ was required to reduce the
sensitivity to target-mass~\cite{pr:d14:1829} and 
higher-twist~\cite{pl:b274:221} contributions which become 
important at low $W^2$.
\end{itemize}
The kinematic range covered by the data input to the fits 
is $6.3 \times 10^{-5}~\leq~ x~\leq~0.65$ and 
$ 2.5~\leq~Q^2~\leq~30,000~\unit{GeV}^2$.

The QCD predictions for the structure functions needed to construct the
reduced cross section
were obtained by solving the DGLAP evolution 
equations at NLO in the \msbar\ 
scheme~\cite{np:b175:27,pl:b97:437,zfp:c11:293}
with the renormalisation and factorisation scales chosen to be $Q^2$.
The DGLAP equations yield the quark and gluon momentum distributions
(and thus the structure functions) at all values of $Q^2$, provided they
are input as functions of $x$ at some input scale, $Q^2_0$.
The input scale was chosen to be 
$Q^2_0 = 7\unit{GeV}^2$; however, there is no particular significance to this value
since backward evolution was performed to fit lower-$Q^2$
data. The choices of the value of $Q^2_0$, the forms 
of the parameterisations of the parton distributions at $Q^2_0$,  
and the cuts on the data to be fitted have all been varied in 
the course of systematic studies (see Section~\ref{sec:model}).

\subsection{Parameterisation of parton distribution functions}
\label{sec:pdf}

The parton distribution functions (PDFs) were parameterised  at
$Q^2_0$ by the form 
\[
   xf(x) = p_1 x^{p_2} (1-x)^{p_3}( 1 + p_5 x)
\]
so that the distributions are either zero, or singular, as $x~\to~0$, and 
tend to zero as $x~\to~1$.
The parton momentum distributions that were parameterised are: 
$u$ valence, $xu_v(x)$; $d$ valence, $xd_v(x)$; 
total sea,
$xS(x)$; gluon, $xg(x)$; and the difference between the $d$ and $u$
contributions to the sea, $x\Delta=x(\bar d-\bar u)$.
The total sea at $Q^2_0$ is made from the flavours up, $xu_{\rm sea}(x)$, down, 
$xd_{\rm sea}(x)$, strange, $xs_{\rm sea}(x)$ and charm, $xc_{\rm sea}(x)$, as follows 
\[
xs_{\rm sea}(x) = 0.2xS(x)
\]
\[
xu_{\rm sea}(x) = 0.4xS(x)-0.5xc_{\rm sea}(x)-x\Delta(x)
\]
\[
xd_{\rm sea}(x) = 0.4xS(x)-0.5xc_{\rm sea}(x)+x\Delta(x)
\]
where the symbols $u_{\rm sea}$, $d_{\rm sea}$, $s_{\rm sea}$, $c_{\rm sea}$ include both
quark and antiquark contributions to the sea for each flavour. The charmed 
sea is generated as described in the next section.
The suppression of the strange sea to $20\%$ of the total sea is
consistent with neutrino-induced dimuon data from CCFR~\cite{zfp:c65:189}.
The fit results are insensitive to this assumption. 

The following parameters were fixed:
\begin{itemize}
\item
$p_1$ for $xu_v$ and $xd_v$ were fixed through the 
number sum-rules and $p_1$ for $xg$ was fixed through the
momentum sum-rule;
\item
$p_2=0.5$ was fixed for both valence distributions, since, after 
the cut $x > 0.1$
on the $xF_3$ data little information on the low-$x$ valence shapes survives.
Allowing these parameters to vary, and varying the value of the low-$x$ cut,
produces values consistent with $0.5$ and has negligible effect on the shapes
of these distributions;
\item
the only free parameter for the
$x\Delta$ distribution is its normalisation, $p_1$, because there is 
insufficient
information on its shape without including E866 Drell-Yan 
data~\cite{prl:80:3715}. 
Thus, $p_2(\Delta)=0.5$, $p_3(\Delta)=p_3({\rm Sea})+2$ were fixed, following 
MRST~\cite{epj:c14:133,npps:79:105,epj:c4:463}, and 
$p_5(\Delta)=0$; the normalisation $p_1(\Delta)$ was found to be
compatible with the measured value of the Gottfried 
sum-rule\cite{pl:b295:159,prl:66:2712}. The fit results are insensitive to
these assumptions;
\item
for the gluon distribution, $p_5$ was set to zero, 
since this choice constrains the high-$x$ gluon to be positive.
Allowing this parameter to vary in the fit produces values which are consistent
with zero.

\end{itemize}
There are thus 11 
free parameters in the ZEUS-S fit, when the strong coupling constant
 is fixed to $\asmz =  0.118$~\cite{epj:c15:1},
and 12 free parameters in the ZEUS-$\alpha_s$ fit.
In the DGLAP evolution equations at NLO, 
$\alpha_s(Q^2)$ is calculated to 2-loop accuracy.
The evolution was performed
with the program QCDNUM~\cite{upub:botje:qcdnum1612}. The evolution
equations were written in terms of quark-flavour singlet and non-singlet
distributions (made from the sea and valence quark distributions) and the 
gluon momentum distribution. These must be convoluted with coefficient 
functions in order to calculate structure functions. The 
coefficient functions are specific to the heavy-quark formalism used, as 
discussed below.

\subsection{Treatment of heavy quarks}
\label{sec:hq}
The treatment of the heavy-quark sea needs careful consideration.
Many early global 
fits~\cite{pr:d47:867,pr:d50:6734,pl:b354:155,epj:c14:133,pl:b387:419,pl:b304:159,pr:d51:4763,pr:d55:1280,epj:c12:375} 
used zero-mass 
variable-flavour-number (ZMVFN) schemes, where, for example, the
charmed quark (of mass $m_c$) is only produced once
$Q^2 > 4 m_c^2$; at larger $Q^2$, the charm distribution is generated by
the splitting $g\rightarrow c \bar c$ using the equations for massless 
partons. This is incorrect at threshold. Other 
authors~\cite{zfp:c67:433,np:b422:37,epj:c5:461} have 
used a fixed-flavour-number (FFN) scheme, in which  a 
$c \bar c$ pair is created by boson-gluon fusion for $W^2 > (2m_c+m_p)^2$, 
(a $W^2$ which may correspond to $Q^2 \ll 4 m_c^2$, if $x$ is small) 
but charm is then treated
as a heavy quark which is dynamically generated for all $Q^2$. There is then
no concept of a charmed parton distribution and thus $\ln(Q^2/m^2_c)$ terms 
remain in the NLO BGF coefficient functions, since they cannot be summed and
absorbed into the definition of the charm distribution. This is
incorrect at high $Q^2$. Recently several 
groups~\cite{zfp:c74:463,epj:c2:287,pr:d50:3102,epj:c1:301,pr:d57:241,pr:d62:096007,pr:d61:096004,epj:c18:547} have tried
to construct general-mass variable-flavour-number schemes which 
behave correctly from threshold to large $Q^2$. 
In this analysis, the scheme of Thorne and 
Roberts~\cite{jp:g25:1307,pl:b421:303,pr:d57:6871,epj:c19:339} (TRVFN) 
has been used to interpolate
between correct threshold and correct large-$Q^2$ behaviour.
The results are compared to those obtained using 
the FFN and ZMVFN schemes in Section~\ref{sec:model}.

\subsection{Definition of $\chi^2$ and treatment of correlated systematic uncertainties}
\label{sec:errors}

The $\chi^2$ minimisation and
the calculation of the covariance matrices were based on
MINUIT~\cite{upub:james:minuit94.1}. The definition of the $\chi^2$ was
\begin{equation}
\chi^2 = \sum_i \frac{\left[ F_i(p,s)-F_i(\rm meas) \right]^2}{(\sigma_{i,\rm stat}^2+\sigma_{i,\rm unc}^2)} + \sum_\lambda s^2_\lambda 
\label{eq:chi2}
\end{equation}
where
\begin{equation}
F_i(p,s) = F_i^{\rm NLOQCD}(p) + 
\sum_{\lambda} s_{\lambda} \Delta^{\rm sys}_{i\lambda}
\label{eq:predmod}
\end{equation}
The symbol $F_i(\rm meas)$ represents a measured data point (structure function
or reduced cross section) and the symbols $\sigma_{i,\rm stat}$ and 
$\sigma_{i,\rm unc}$ represent its error from statistical and uncorrelated 
systematic uncertainties, 
respectively. The symbol $F_i^{\rm NLOQCD}(p)$ represents the prediction 
from NLO QCD in terms of the theoretical parameters $p$
(PDF parameters and $\asmz$). This prediction is modified
to include the effect of the correlated systematic uncertainties as shown in
Eq.~(\ref{eq:predmod}). 
The one-standard-deviation systematic uncertainty 
on data point $i$ due to source $\lambda$ is
referred to as $\Delta^{\rm sys}_{i\lambda}$ and 
the parameters $s_\lambda$ represent independent Gaussian random variables 
with zero mean and unit variance for each source of
 systematic uncertainty. These
parameters $s_\lambda$ were fixed to zero to obtain the 
central values of the theoretical parameters,
but they were allowed  
to vary for the error analysis, such that in addition to the usual
Hessian matrix, $M_{jk}$, given by
\[
M_{jk}= \frac{1}{2}\frac{\partial^2 \chi^2}{\partial p_j \partial p_k},
\]
which is evaluated with respect to the theoretical parameters,
a second Hessian matrix, $C_{j\lambda}$, given by
\[
C_{j\lambda} = \frac{1}{2}\frac{\partial^2 \chi^2}{\partial p_j \partial s_{\lambda}}
\]
was evaluated. The systematic covariance matrix is then given by
$V^{ps}= M^{-1} C C^T M^{-1}$~\cite{pasc:lal:9505} and the total
covariance matrix by  
$V^{\rm tot} = V^p +V^{ps}$, where $V^p=M^{-1}$. 
Then the uncertainty on any distribution $F$ may be calculated from
\[
<\Delta F^2>=\sum_j \sum_k \frac{\partial F}{\partial p_j} V_{jk} \frac{\partial F}{\partial p_k} 
\]
by substituting $V^p$, $V^{ps}$ or $V^{\rm tot}$ for $V$, to obtain the
statistical (and uncorrelated systematic), correlated systematic 
or total experimental error band, respectively. 
This method of accounting for systematic uncertainties is equivalent, to first
order, to the 
`offset method', in which each $s_\lambda$ is varied by its 
assumed uncertainty ($\pm1$), 
a new fit is performed for each of these variations,
and the resulting deviations of the theoretical parameters 
from their  central  values are added in 
quadrature~\cite{jp:g28:779}. 
Either of these methods of treating systematic uncertainties results in more
conservative error estimates than alternative methods discussed 
in the Appendix.

The normalisations of the data sets were taken as published,  
apart from BCDMS data, which were scaled down~\cite{mv:dphpe:9007,epj:c14:133,npps:79:105,epj:c4:463,pl:b387:419} by $2\%$. However, the 
normalisation uncertainties were included among the correlated systematic 
uncertainties. In total, 71 independent sources of systematic uncertainty
were included (see Table~\ref{tab:tabchi}). 

\section{Fit results, theoretical and model uncertainties 
and the extraction of $\mathbf{\asmz}$}
\label{sec:standard}

\subsection{Fit quality and fit predictions}
\label{sec:fitqual}

\begin{figure}[!p]
\includegraphics[scale=.65]{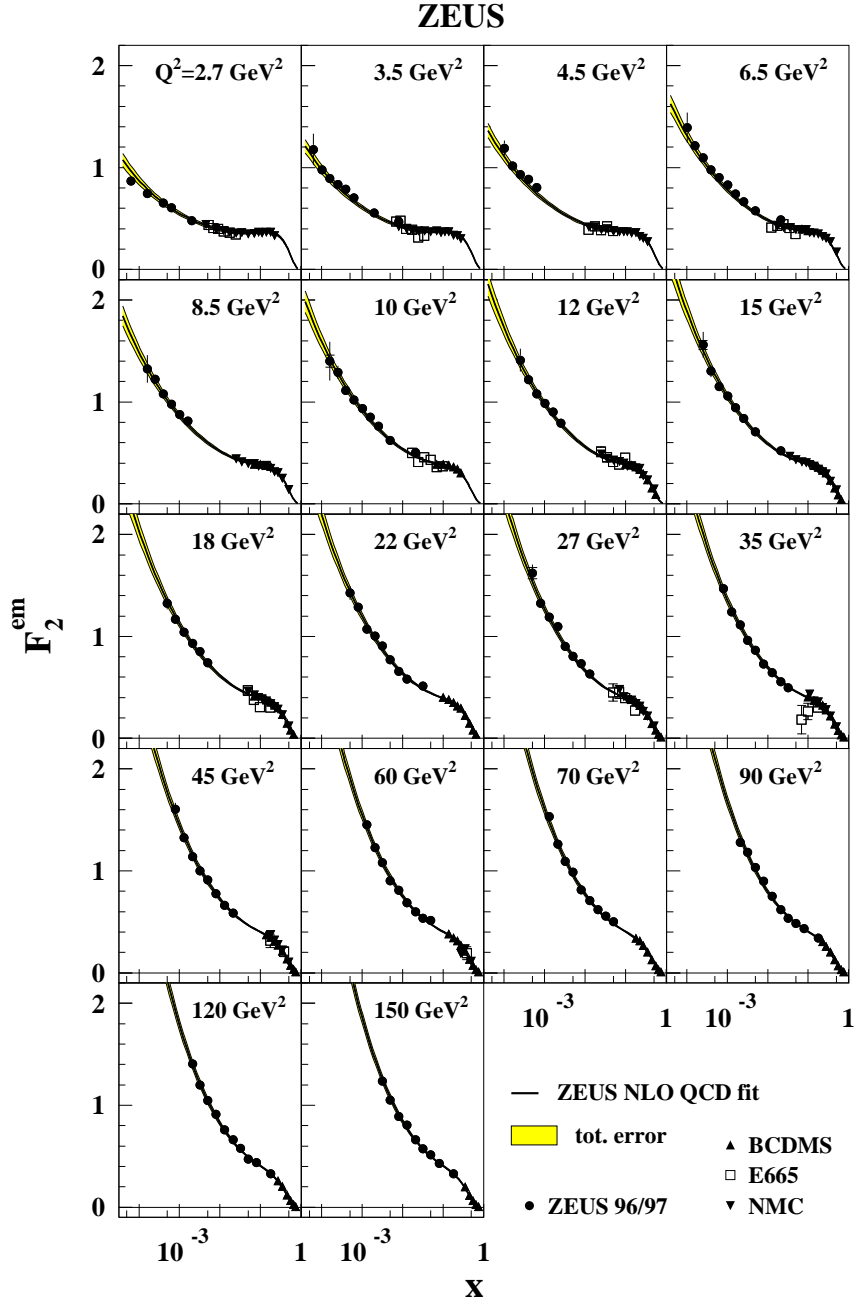}
\caption{The ZEUS-S NLO QCD fit compared to ZEUS 96/97  and proton 
fixed-target $F_2$ data. The error bands of the fit 
represent the total experimental
uncertainty from both correlated and uncorrelated sources. 
}
\label{fig:f2lowall}
\end{figure}

\begin{figure}[!p]
\includegraphics[scale=.65]{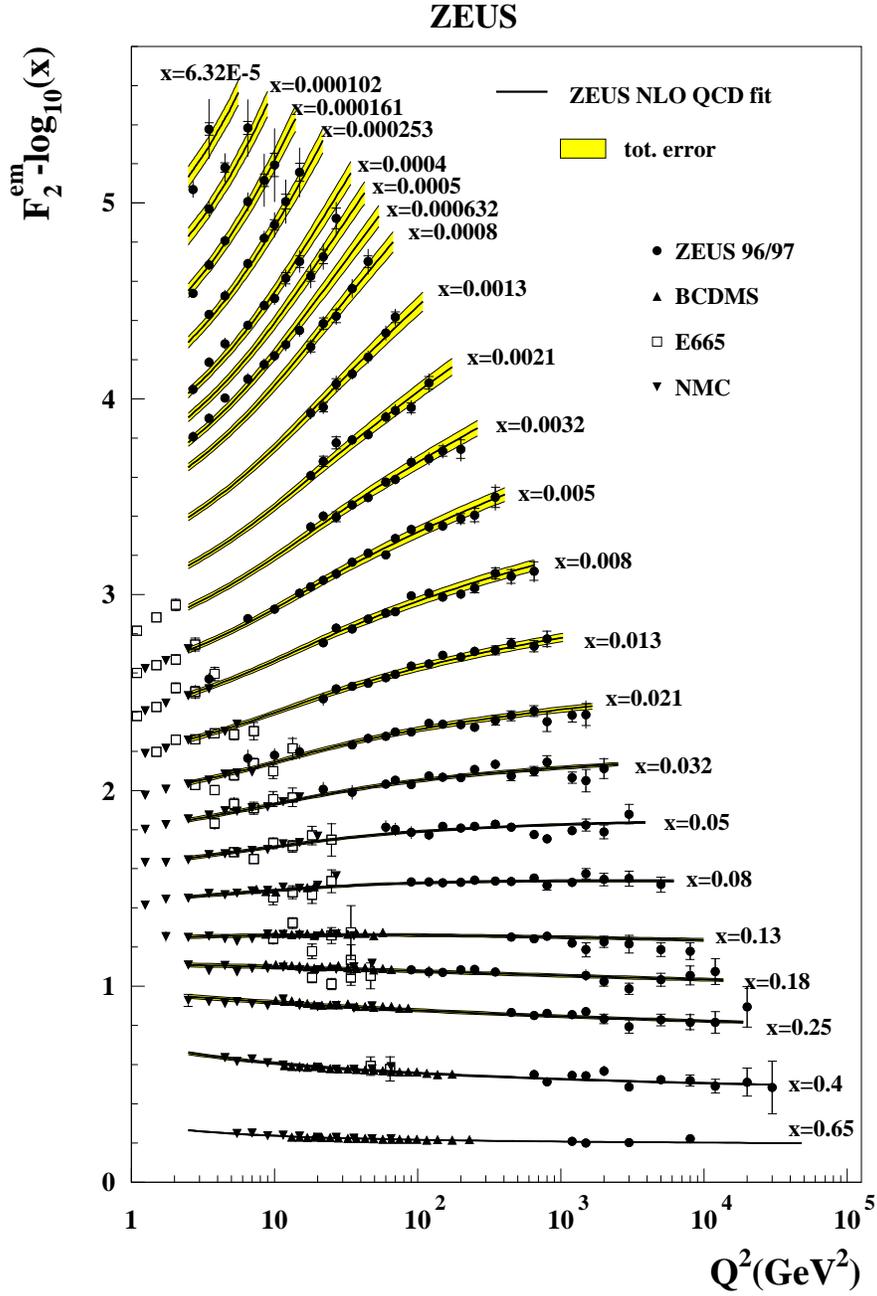}
\caption{The ZEUS-S NLO QCD fit compared to 
ZEUS 96/97 and proton fixed-target 
$F_2$ data. The error bands are defined in the caption to Fig.~\ref{fig:f2lowall}.}
\label{fig:scalvio}
\end{figure}

\begin{figure}[!p]
\includegraphics[scale=.65]{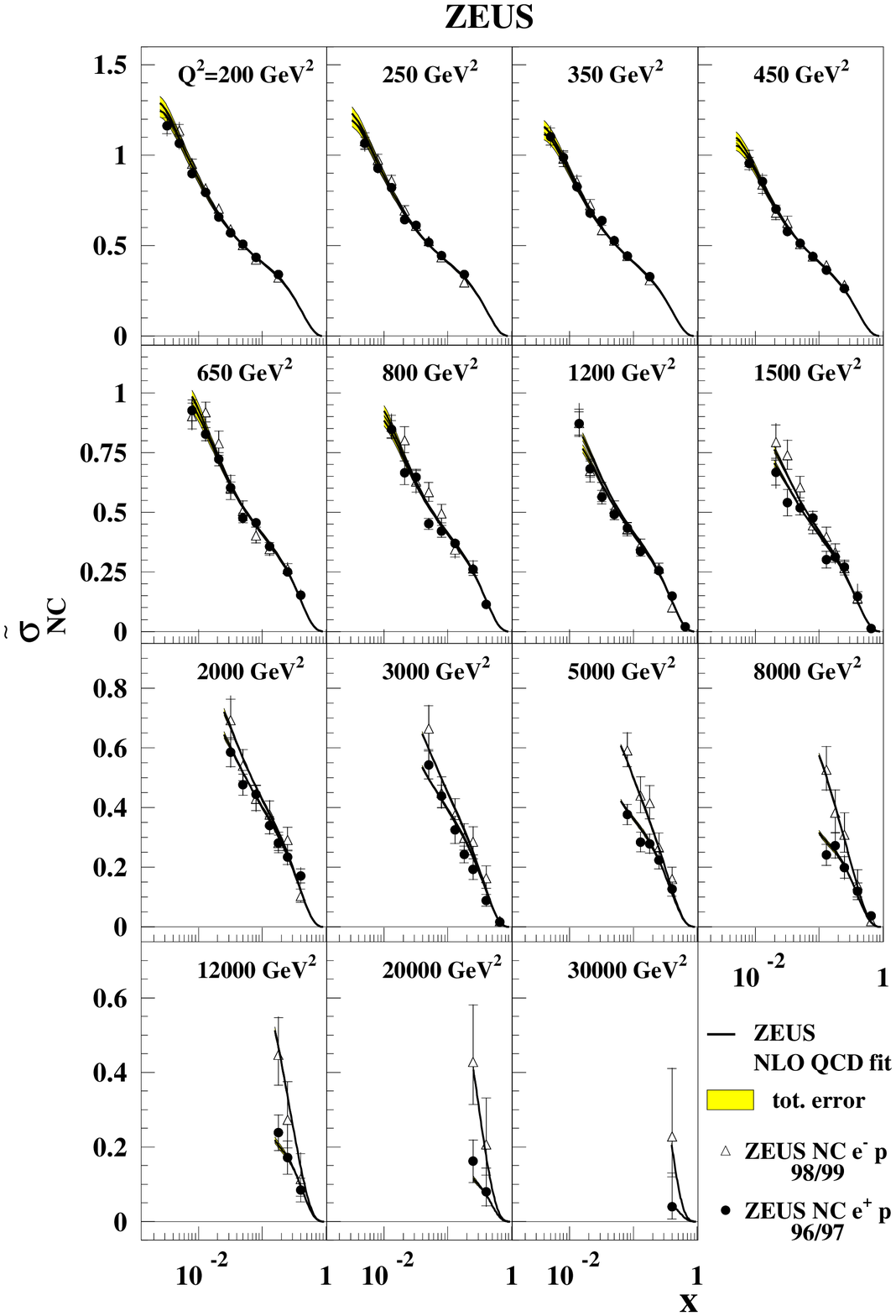}
\caption{The ZEUS-S NLO QCD fit compared to the ZEUS high-$Q^2$ $e^+p$ 
and $e^-p$
 neutral current reduced cross sections. The error bands are defined in the caption to
Fig.~\ref{fig:f2lowall}. Note that the $e^+p$ data were taken 
at $\surd s = 300 \unit{GeV}$, 
whereas the $e^-p$ data were taken at $\surd s = 318 \unit{GeV}$.}
\label{fig:ncelpo}
\end{figure}

\begin{figure}[!p]
\includegraphics[scale=.65]{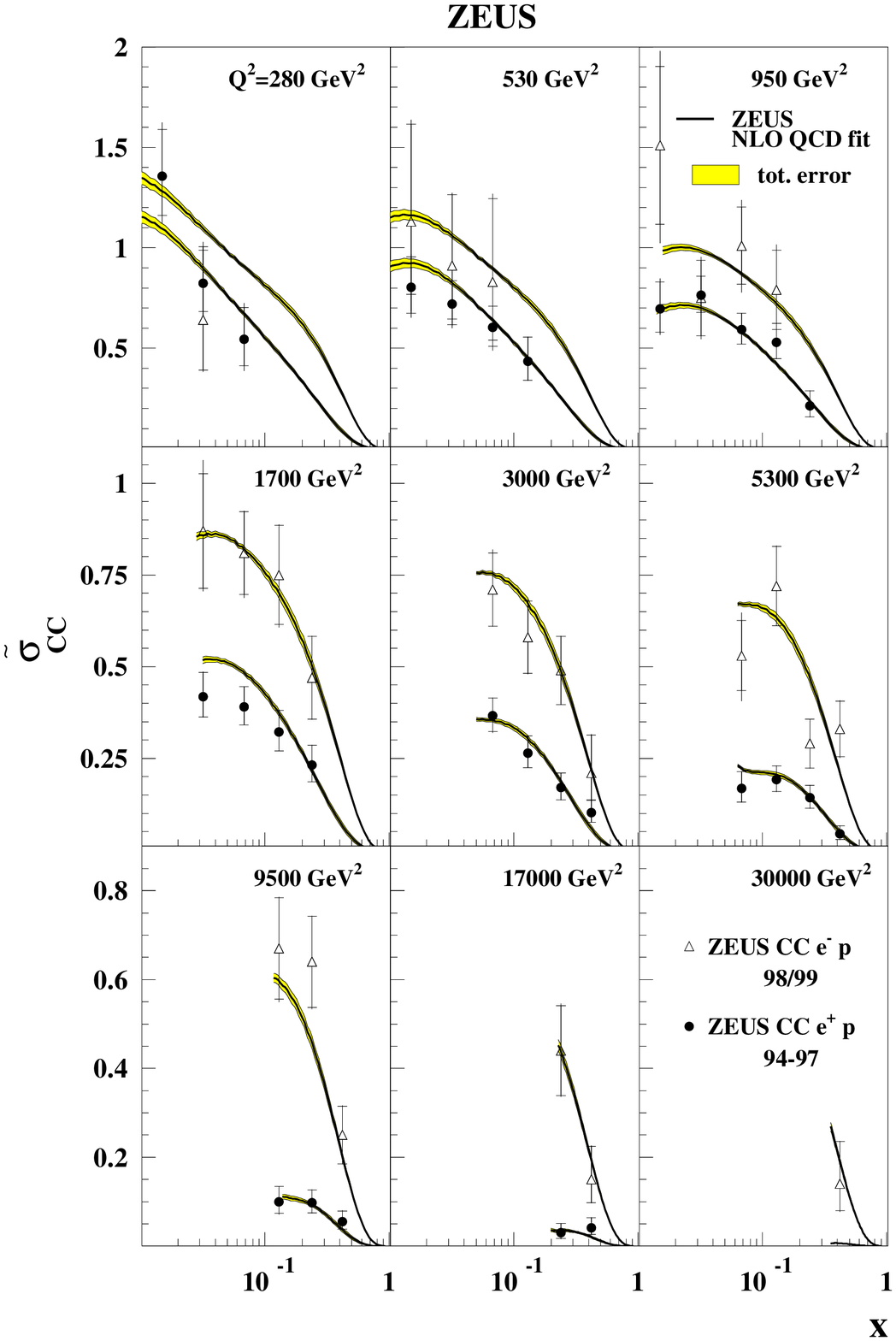}
\caption{The ZEUS-S NLO QCD fit compared to the ZEUS high-$Q^2$ $e^+p$ 
and $e^-p$
 charged current reduced cross sections. The error bands are defined in the caption to
 Fig.~\ref{fig:f2lowall}. Note that the $e^+p$ data were taken at 
$\surd s = 300 \unit{GeV}$, 
whereas the $e^-p$ data were taken at $\surd s = 318 \unit{GeV}$.}
\label{fig:ccelpo}
\end{figure}

The ZEUS-S fit, with $\asmz = 0.118$, is shown in 
Figs.~\ref{fig:f2lowall}-\ref{fig:ccelpo}. In Fig.~\ref{fig:f2lowall}, 
the fit prediction for $F_2$ is shown compared to the ZEUS and proton 
fixed-target data as a function
of $x$ at low $Q^2$. In Fig.~\ref{fig:scalvio}, this comparison is made
as a function of $Q^2$ for $x$ values in the range 
$6.3 \times 10^{-5} < x < 0.65$. 
For the fixed-target data, only the $\gamma$-exchange process contributes to
$F_2$, whereas, at high-$Q^2$, there are also contributions 
from $Z^0$ exchange and $\gamma/Z^0$ interference. Thus, for comparability
with the fixed-target results, the ZEUS data
shown in these figures represent only that part 
of $F_2$ due to $\gamma$ exchange, as denoted by the symbol $F_2^{em}$.
The fit gives an excellent description of the data.

The goodness of fit cannot be judged from the $\chi^2$
calculated from statistical and uncorrelated systematic errors alone.
Re-evaluating the $\chi^2$ for the parameters 
resulting from the ZEUS-S fit by
adding the statistical, uncorrelated and correlated systematic errors 
in quadrature
gives a total $\chi^2$ per data point  of 0.95 for 1263 data points 
and 11 free parameters.
The $\chi^2$ per data point for individual data sets 
calculated in the same way are listed in Table~\ref{tab:tabchi}. 

In Fig.~\ref{fig:ncelpo}, the fit is compared to the ZEUS high-$Q^2$ 
neutral current $e^+p$ data. This figure also shows 
predictions for the neutral current $e^-p$
data~\cite{desy:02:113}, which
were not included in the fit. The charged current $e^+p$~\cite{epj:c12:411} and
$e^-p$~\cite{pl:b539:197} data (which were also not
included in the fit) are compared to the fit prediction 
in  Fig.~\ref{fig:ccelpo}.
These high-$Q^2$ data are very well described by the fit.

\subsection{Parton distribution functions and $F_L$}
\label{sec:pdfres}
\begin{figure}[t]
\vspace{-4.0cm}
\centerline{\psfig{figure=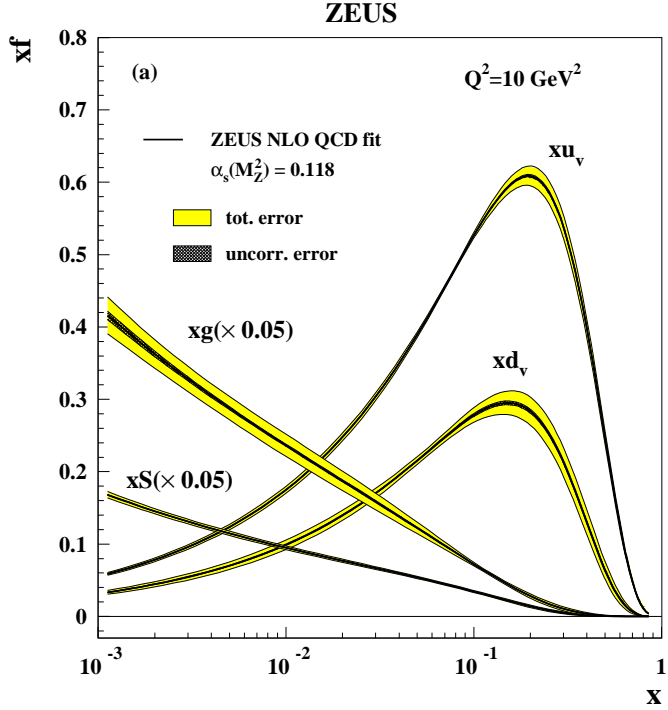,width=11cm}}
\vspace{-4.0cm}
\centerline{\psfig{figure=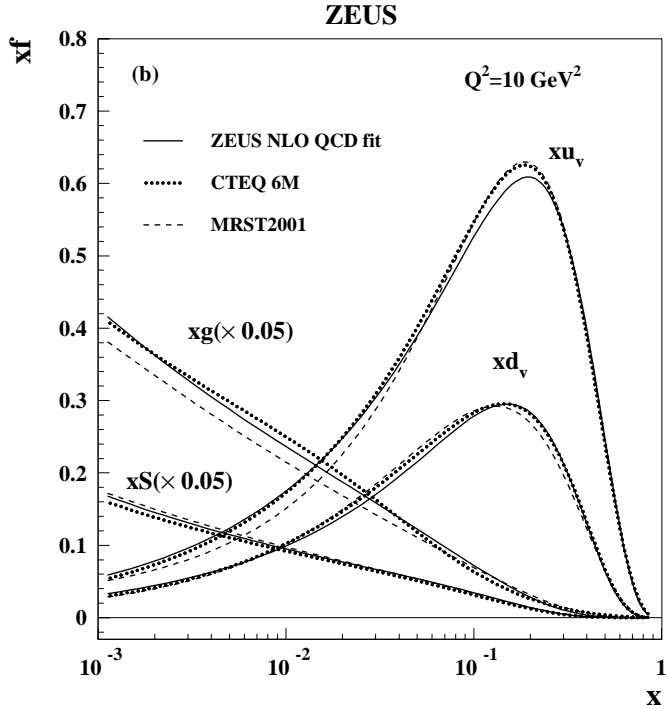,width=11cm}}
\vspace{-2.0cm}
\caption[]{(a)
The gluon, sea, $u$ and $d$ valence distributions extracted from 
the standard ZEUS-S NLO QCD fit at $Q^2 = 10\unit{GeV}^2$. 
The error bands 
in this figure show the uncertainty from statistical and
other uncorrelated sources separately from the total uncertainty including
correlated systematic uncertainties. (b) 
The gluon, sea, $u$ and $d$ valence distributions extracted from 
the ZEUS-S NLO QCD fit at $Q^2 = 10\unit{GeV}^2$, compared to those
extracted from the fits MRST2001~\cite{epj:c23:73} and 
CTEQ6~\cite{hep-ph-0201195}.}
\label{fig:erd}
\end{figure}

The PDF parameters extracted from the ZEUS-S fit at $Q^2_0 = 7\unit{GeV}^2$ 
are given in Table~\ref{tab:param} and
the corresponding parton distributions at $Q^2 = 10\unit{GeV}^2$ are shown in 
Fig.~\ref{fig:erd}(a). 
The precision of these distributions is considerably improved 
in comparison to a fit~\cite{epj:c14:285} using earlier ZEUS data.
The total error band is dominated by systematic uncertainties.
In Fig.~\ref{fig:erd}(b), the ZEUS 
parton distributions are compared to the 
latest distributions from MRST~\cite{epj:c23:73} and 
CTEQ~\cite{hep-ph-0201195}. 
The differences between these sets of 
parton distributions are compatible with the size of the error bands 
on the ZEUS parton distributions.

The PDFs extracted from the ZEUS-S fit are now considered in more 
detail. In these distributions, the contribution to the error bands coming from
variation of  
$\asmz$ will be indicated in addition to the contributions of correlated 
and uncorrelated experimental uncertainties. This additional 
uncertainty has been taken into account with full correlations by allowing 
$\asmz$ to be a parameter of the ZEUS-$\alpha_s$ fit 
(see Section~\ref{sec:alphas}).

\begin{figure}[!p]
\includegraphics[scale=.75]{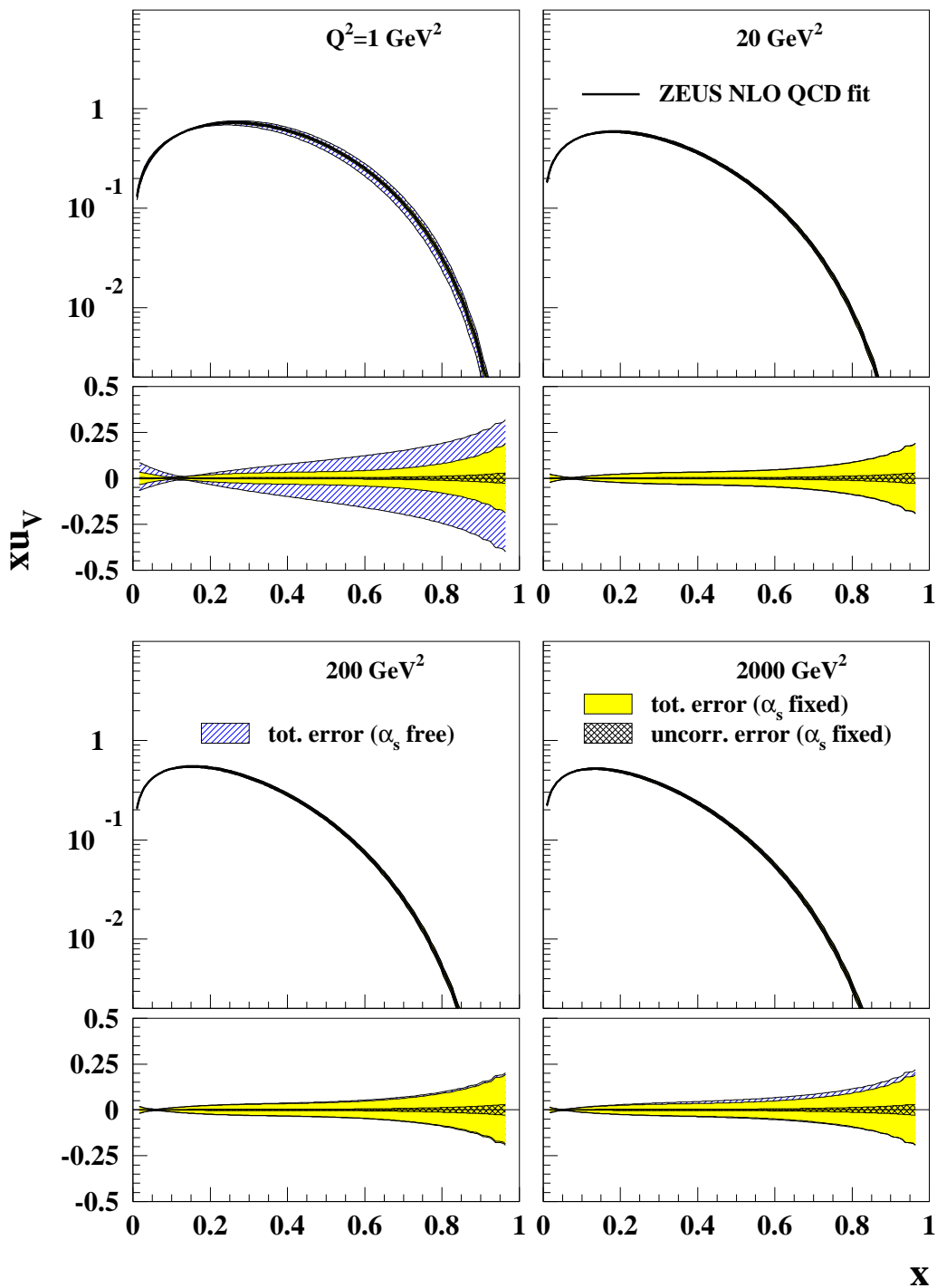}
\caption{The $xu_v$ distribution  
from the ZEUS-S NLO QCD fit. 
The cross-hatched error bands show the statistical and 
uncorrelated systematic uncertainty, the grey error bands show the total 
experimental uncertainty including correlated systematic uncertainties 
(both evaluated from the ZEUS-S fit) and the hatched error bands show 
the additional uncertainty coming from variation 
of the strong coupling constant $\asmz$ 
(evaluated from the ZEUS-$\alpha_s$ fit). The uncertainties on these
distributions are shown beneath each distribution as fractional differences
from the central value.
}
\label{fig:uv}
\end{figure}

\begin{figure}[!p]
\includegraphics[scale=.75]{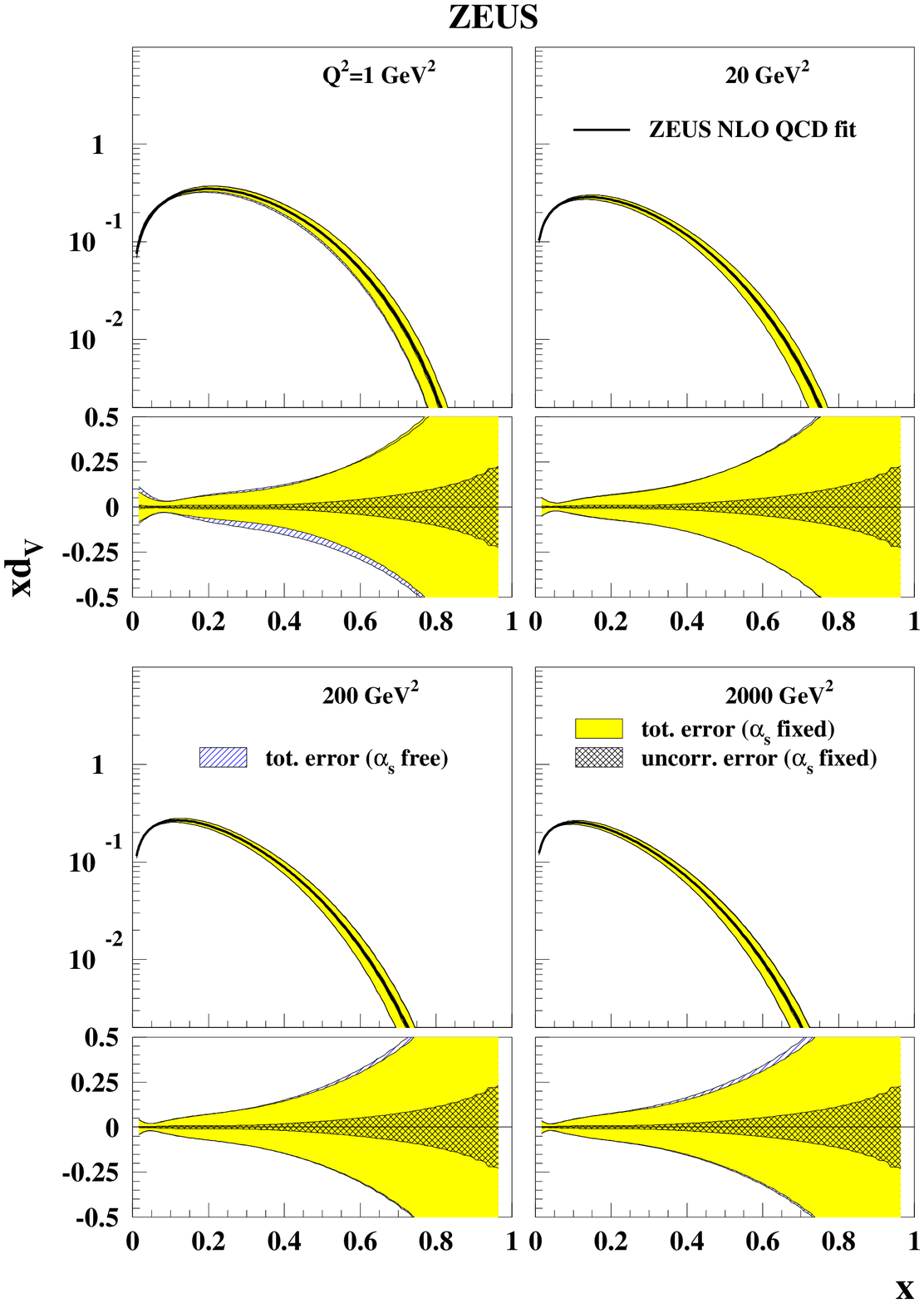}
\caption{The $xd_v$ distribution  
from the ZEUS-S NLO QCD fit. The error bands are defined in the caption to 
Fig.~\ref{fig:uv}.
The uncertainties on these
distributions are shown beneath each distribution as fractional differences
from the central value.
}
\label{fig:dv}
\end{figure}

The valence distributions $xu_v$ and $xd_v$ extracted from the fit are
shown for several different $Q^2$ values in 
Figs.~\ref{fig:uv} and \ref{fig:dv}. 
The abscissa is linear and the ordinate logarithmic
to illustrate the high-$x$ behaviour of these valence distributions, 
where they are constrained by the fixed-target data. The distributions 
for $Q^2 = 1\unit{GeV}^2$ are obtained by backward extrapolation.
The uncertainty is 
shown beneath each distribution in terms of the fractional 
differences from the central value. The $u$-valence
distribution is much better determined than the $d$-valence distribution, 
since structure-function data from fixed-target experiments 
are dominated by the $u$ quark. 

\begin{figure}[!p]
\includegraphics[scale=.65]{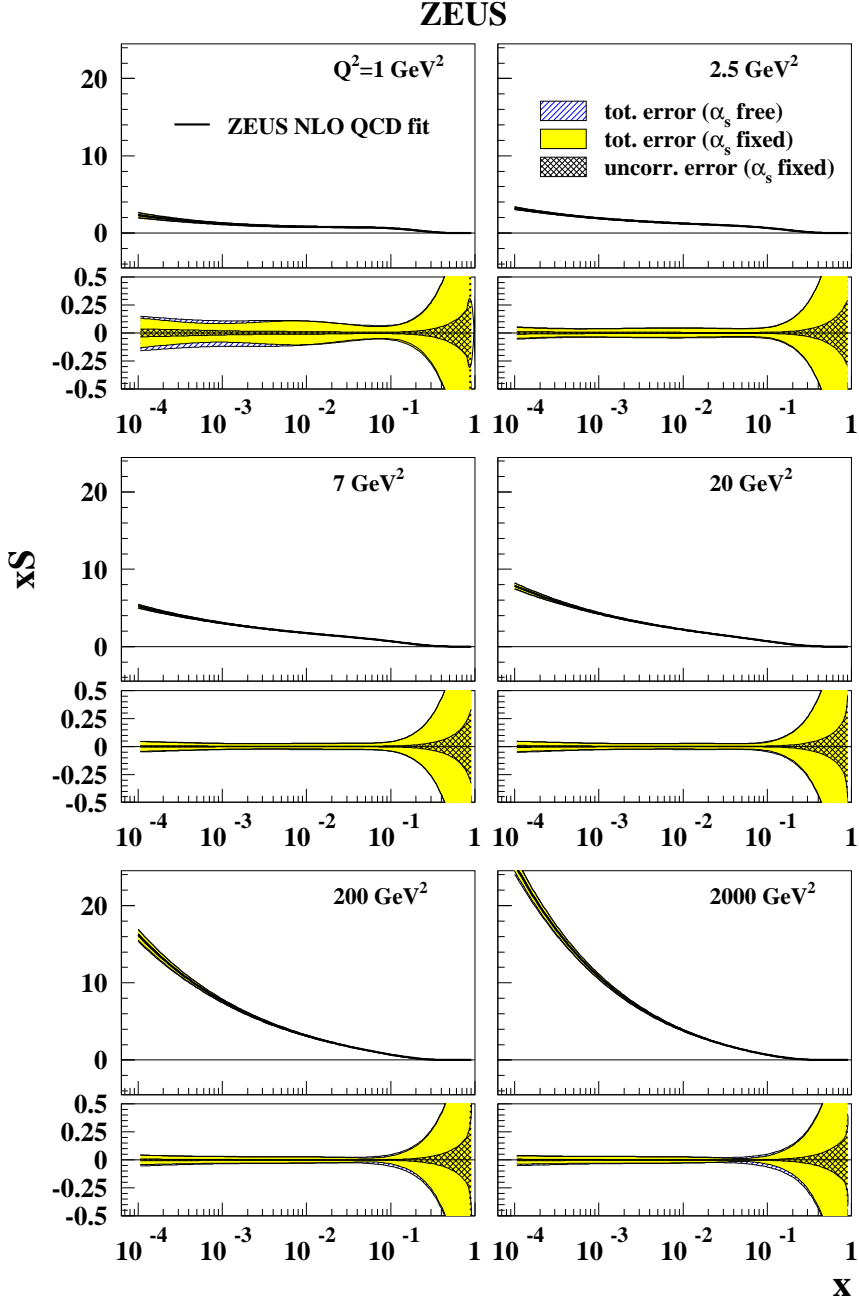}
\caption{The sea distribution from the ZEUS-S NLO QCD fit for various
$Q^2$ values. The error bands are defined in the caption to Fig.~\ref{fig:uv}.
The uncertainties on these
distributions are shown beneath each distribution as fractional 
differences from the central value.
}
\label{fig:sea}
\end{figure}

The extracted sea distribution and its uncertainty is shown for several 
$Q^2$ values in Fig.~\ref{fig:sea}. 
The uncertainty in these distributions is less than
$\sim 5\%$ for $Q^2 \gtrsim 2.5\unit{GeV}^2$ and  $10^{-4} < x < 10^ {-1}$,
but considerable uncertainty remains for
$x > 0.1$. The sea distribution rises at small $x$, 
even at $Q^2 = 1\unit{GeV}^2$. 

\begin{figure}[!p]
\includegraphics[scale=.65]{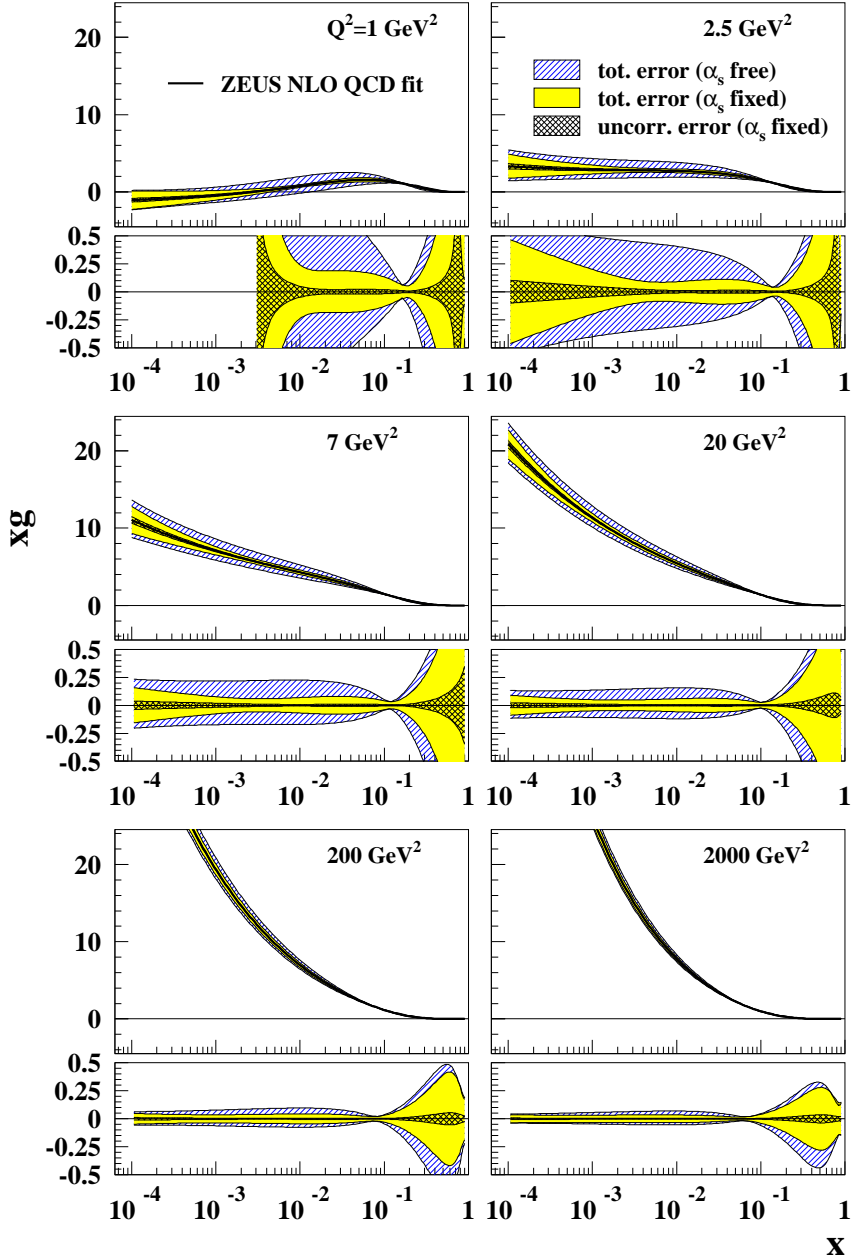}
\caption{The gluon distribution from the ZEUS-S NLO QCD fit for various
$Q^2$ values.
The error bands are defined in the caption to Fig.~\ref{fig:uv}. The uncertainties on these
distributions are shown beneath each distribution as fractional 
differences from the central value. Note that this  
uncertainty is not shown
when the central value of the gluon distribution becomes negative.
}
\label{fig:gbig}
\end{figure}

The corresponding gluon distribution and its uncertainty 
is shown for several $Q^2$ values in Fig.~\ref{fig:gbig}.
The general shape of the error bands, with a narrowing at $x \sim 0.1$,
is a consequence of the momentum sum rule.
The gluon distribution is 
determined to within $\sim 10\%$ for $Q^2 > 20\unit{GeV}^2$ and  
$10^{-4} < x < 10^ {-1}$; and its uncertainty
decreases as $Q^2$ increases. Considerable uncertainty remains for
$x > 0.1$. The distribution rises steeply at low $x$ for 
$Q^2 \gtrsim 5\unit{GeV}^2$; however,
at lower $Q^2$, the low-$x$ gluon shape is flatter. When
the fit is extrapolated back to $Q^2 = 1\unit{GeV}^2$, the shape
becomes valence-like and tends to become negative at the lowest $x$, 
although remaining consistent with zero. 

\begin{figure}[!p]
\includegraphics[scale=.65]{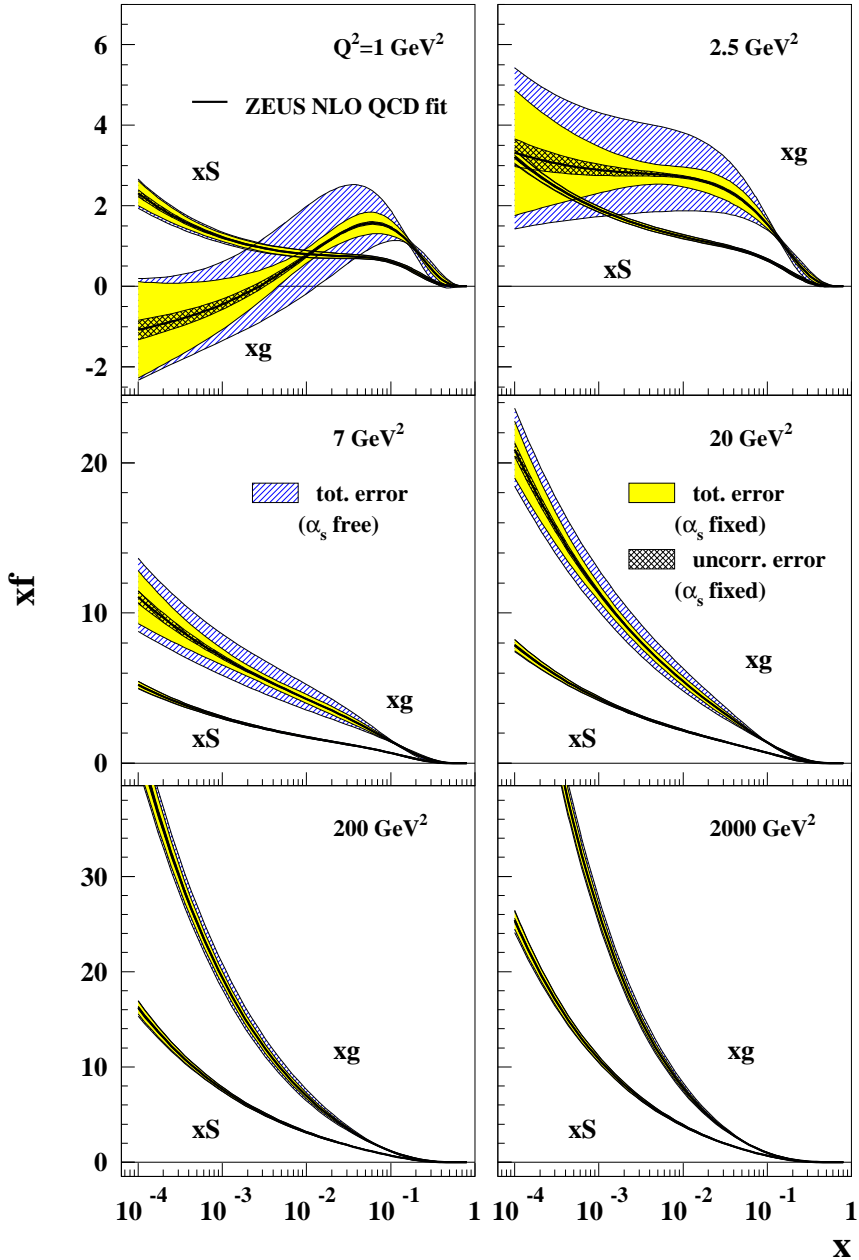}
\caption{Comparison of the  gluon and sea distributions from the 
ZEUS-S NLO QCD fit for various $Q^2$ values.
The error bands are defined in the caption to Fig.~\ref{fig:uv}. 
}
\label{fig:glusea}
\end{figure}

The shapes of the gluon and the 
sea distributions are compared in Fig.~\ref{fig:glusea}. For 
$Q^2 \gtrsim 5\unit{GeV}^2$, the gluon density becomes much larger than 
the sea density, but for lower $Q^2$ the sea density continues to rise
at low $x$, whereas the gluon density is suppressed.
The present analysis 
shows this contrasting behaviour of the low-$x$, low-$Q^2$ gluon and sea
distributions even more clearly than the previous study of earlier 
ZEUS data~\cite{epj:c7:609}.

\begin{figure}[t]
\vspace{-2.5cm}
\centerline{\psfig{figure=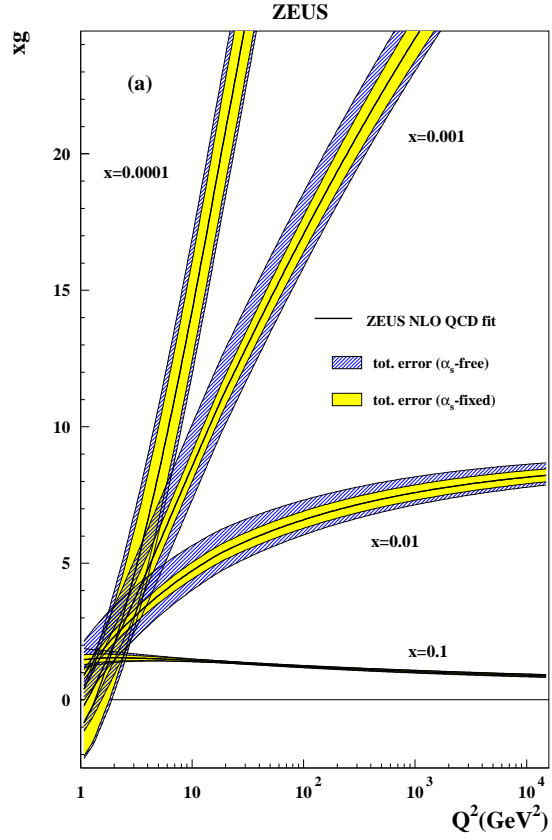,width=9cm}}
\centerline{\psfig{figure=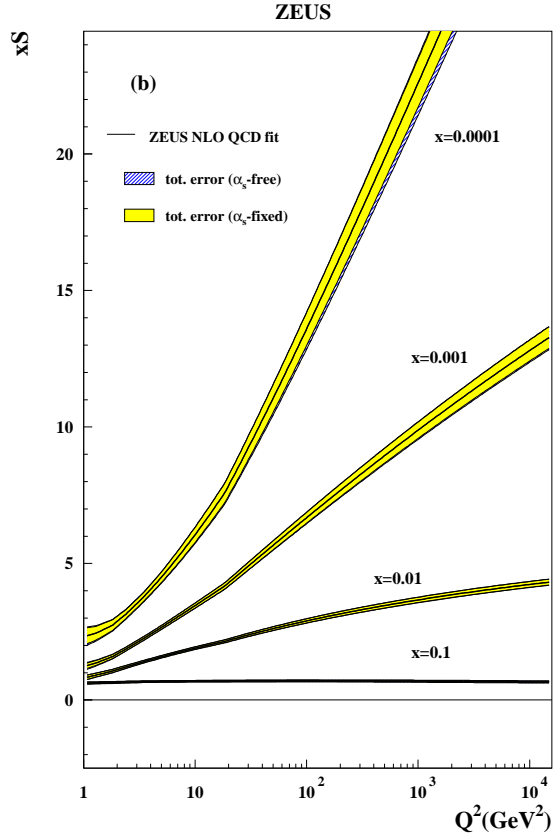,width=9cm}}
\caption[]{(a) The gluon distribution from the ZEUS-S NLO QCD fit
as a function of $Q^2$ for fixed $x$ values
 (b) The sea distribution from 
the ZEUS-S NLO QCD fit
as a function of $Q^2$ for fixed $x$ values.
The error bands are defined in the caption to Fig.~\ref{fig:uv}.
}
\label{fig:scal}
\end{figure}

It is also interesting to compare the behaviour of the gluon and the sea
NLO densities as a function of $Q^2$ for fixed $x$ values. This is shown in 
Fig.~\ref{fig:scal}. The scaling 
violation of the gluon distribution at small $x$ is striking, reflecting the
singular behaviour of the $P_{gq}$ and $P_{gg}$ splitting functions as 
$x \to 0$.

\begin{figure}[!p]
\vspace{-3.0cm}
\includegraphics[scale=.75]{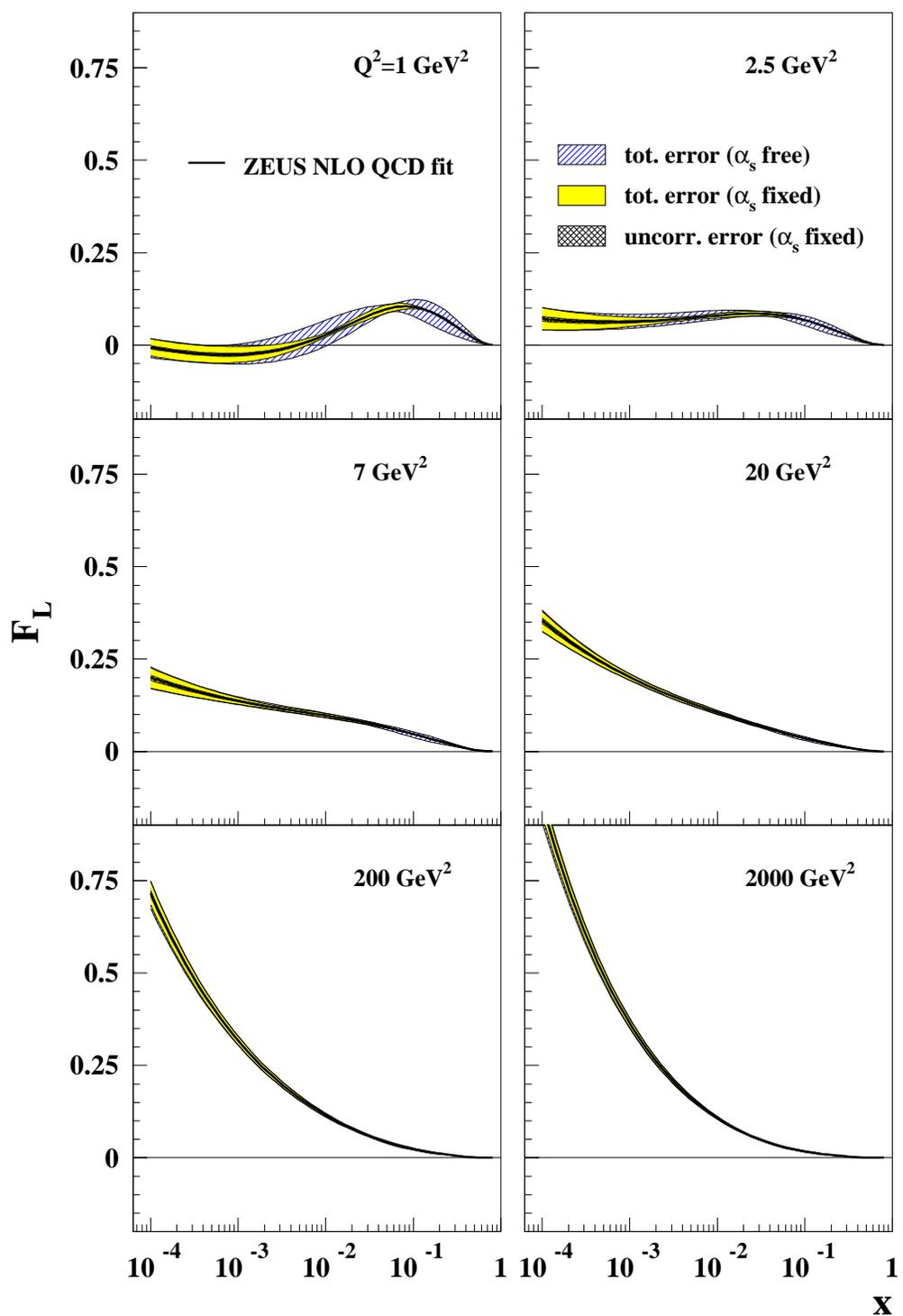}
\caption{The longitudinal structure function $F_L$
 from the ZEUS-S NLO QCD fit. The error bands are defined in the caption to 
Fig.~\ref{fig:uv}.}
\label{fig:fl}
\end{figure}

The tendency of the gluon distribution to become negative at low $x$ and 
low $Q^2$ could be a signal that 
NLO QCD is inadequate in this kinematic region. However, the only physical
requirement is that the structure functions calculated from
the parton distributions are positive. Thus it is important to
investigate the fit prediction for $F_L$,
the structure function most closely related to the gluon~\cite{zfp:c39:281}. 
This is shown in Fig.~\ref{fig:fl}. It exhibits similar 
features to the gluon. This will be discussed further in 
Section~\ref{sec:loq2qcd}. 

\subsection{The extraction of $\mathbf{\asmz}$}
\label{sec:alphas}

In the evolution of singlet quark distributions at intermediate $x$ 
($0.01 < x <0.3$),
the value of $\asmz$ and the gluon shape are strongly correlated through 
the DGLAP equations, since an increase in $\asmz$ 
can be compensated by a harder gluon distribution.
This has restricted the precision of determinations of $\asmz$ from 
NLO DGLAP fits to DIS data in the past. 
However, at small $x$ ($x<0.01$) this correlation is weakened, since the gluon
then drives the behaviour of $F_2$ as well as that of $dF_2/d\ln(Q^2)$.
Thus, precision low-$x$ data can be used to make a simultaneous fit
for $\asmz$ and the PDF parameters. 
In the ZEUS-$\alpha_s$ fit,  $\asmz$ was left free, leading to
\begin{equation}
\asmz = 0.1166 \pm 0.0008(\rm uncorr.) \pm 0.0032 (\rm corr.) \pm 0.0036(\rm norm.),
\label{eq:alphas}
\end{equation}
where the three uncertainties arise from the following: 
statistical and other uncorrelated sources; correlated systematic sources from
all the contributing experiments except that from their normalisations; the
contribution from the latter normalisations.

The difference between this value of $\asmz$ 
and the value $0.118$ used in the ZEUS-S fit 
does not produce any significant shifts in the PDF
parameters as compared to those determined in the ZEUS-S fit. 
However, the correlation between $\asmz$ and the PDF parameters does increase 
their experimental uncertainties, particularly that of the gluon, as
illustrated in Figs.~\ref{fig:uv}-\ref{fig:fl}. 

\subsection{Model uncertainties}
\label{sec:model}

Sources of model uncertainty within the theoretical framework 
of leading-twist NLO QCD are now considered.
The sensitivity of the results to the variation of input 
assumptions has been quantified in terms of the resulting variation in $\asmz$,
since it is the most sensitive parameter. 

Table~\ref{tab:model}
summarizes the effect of varying the value of $Q^2_0$, and the minimum $Q^2$, 
 $x$ and $W^2$ ($Q^2_{\rm min}$, $x_{\rm min}$, $W^2_{\rm min}$) 
of data entering the ZEUS-$\alpha_s$ fit, 
in terms of the shift in the central
value of $\asmz$. These variations  produce only a small model uncertainty
in $\asmz$ and in the PDF parameters.

It is also necessary to 
consider varying the form of the input PDF parameterisations.
Variation in the gluon parameterisation produces the most significant effects
since it is least well known. 
Allowing the high-$x$ gluon to take a more complex form, with 
$ p_5 \not= 0$,
resulted in a shift of $\Delta \asmz=+0.0002$. Extending
the form of the parameterisation from $(1 + p_5 x)$ to 
$(1 + p_4 \sqrt x + p_5 x)$ for both the gluon and the other 
parton distributions resulted in a shift of $\Delta \asmz=+0.0008$. Allowing
$p_2$ to be free for the valence distributions had no further 
effect on the value of $\asmz$. 
Finally, information from Tevatron high-$E_T$ jet 
production~\cite{prl:86:1707,pr:d64:012001} was used to
constrain the high-$x$ gluon~\cite{epj:c23:73}. 
The corresponding shift
in the central value of $\asmz$ was  $\Delta \asmz = +0.0006$ and the shape
of the gluon was shifted to be harder at high $x$. However,
these shifts are well within the error estimates for both $\asmz$ and the 
gluon PDF parameters. 

A further significant choice is that of the heavy-quark production scheme. 
Repeating the fit using the FFN scheme or the ZMVFN scheme produced
 shifts of $\Delta \asmz = \pm 0.0010$. 
Variation of the heavy-quark mass within
the FFN and TRVFN schemes produced smaller shifts.
The choice of the heavy-quark scheme also affects the shape of the gluon, such
that the FFN scheme gluon is steeper at small $x$ than the ZMVFN scheme gluon,
with the TRVFN gluon in between. The size of these shifts 
is well within the error estimates of the gluon PDF parameters.

Thus, the total model uncertainty on $\asmz$ is considerably smaller than 
the errors from correlated systematic and normalisation uncertainties and 
leads to
\[
\asmz = 0.1166 \pm 0.0008(\rm uncorr.) \pm 0.0032(\rm corr.) \pm 0.0036(\rm norm.) \pm 0.0018(model).
\]

The PDF parameters are much less sensitive to 
the model assumptions than is $\asmz$. It follows that 
the error bands illustrated on the 
parton densities in Figs.~\ref{fig:erd}-\ref{fig:scal} represent 
reasonable estimates of the total uncertainties.

\subsection{Uncertainties in the theoretical framework}
\label{sec:theoryunc}

While the uncertainty within the theoretical framework of leading-twist NLOQCD
is rather well defined, it is much more difficult to decide on the uncertainty
caused by reasonable variations in the framework. In this Section, 
two variations on the framework are estimated; the treatment of higher-twist 
terms and the choice of the renormalisation and factorisation scales, which
gives an estimate of the importance of the higher-order terms in the pQCD
expansion.

The analysis was performed at leading twist and accordingly
a hard $W^2$ cut was made to remove the region where higher-twist effects are
known to be important. 
In order to evaluate if there are residual effects of higher twist at such
large $W^2$, this cut was lowered to $W^2 > 4\unit{GeV}^2$ and the 
SLAC data~\cite{pl:b282:475} were included\footnote{Note that the $\chi^2$ 
for these
data must be calculated by adding statistical and systematic errors in
quadrature, since information on correlated point-to-point systematic 
uncertainties is not available.}. A fit
in which the leading-twist predictions for $F_2$ were modified by a factor
$(1+h_i/Q^2)$ was then performed, 
where $h_i$, $i=1,10$, are parameters determined in ten bins of 
$x$~\cite{pl:b443:301}.
This modification was not intended to provide a thorough study of
the higher-twist effects themselves, but only as an estimate of the 
uncertainty 
introduced by neglecting them. Hence, a simple form of the higher-twist 
contribution was used, in which
$xF_3$ was not modified and the higher-twist terms for deuterium and proton 
targets were assumed to be the same.
The contribution of higher twist was found to be negligible for $x < 0.005$, 
small and negative for $ 0.005 < x < 0.5$ and large and positive
for $x > 0.5$, where target-mass effects are important. 
Having determined the $h_i$ parameters in this fit, these parameters were 
fixed and a fit was performed with the usual hard $W^2$ cut 
(excluding SLAC data).
This produced a shift of $\Delta \asmz = -0.0032$. 

Variation of the renormalisation and factorisation scales used in the
fit was also considered. 
The choice of $Q^2$ for these scales is conventional in 
the inclusive DIS process, and their variation is used
as a crude way of estimating the importance of higher-order
terms~\cite{epj:c18:117,pr:d64:074005,hep-ph-0102151,pl:b474:372}.
These scales were varied from $Q^2/2 \rightarrow 2Q^2$, 
independently and simultaneously.
This produced shifts $\Delta \asmz \sim \pm 0.004$, mostly from the
change in renormalisation scale. 
The result of making larger scale changes, such as $Q^2/4 \rightarrow 4Q^2$, 
is not presented because such large scale changes
produce fits with much larger $\chi^2$, which are unacceptable
according to the `hypothesis testing' criterion (see Appendix).
It is unclear that such arbitrary scale changes give any reasonable 
estimate of the importance of higher-order terms~\cite{epj:c23:73}. 
Several groups~\cite{pl:b531:216,pl:b519:57,np:b611:447} have compared
NLO and approximate NNLO analyses. 
The change in $\asmz$ obtained in these studies is in the range 
$ -0.0035 < \Delta \asmz < -0.001$. 

The uncertainties discussed in this Section are rather large. However, since 
these investigations are far from exhaustive and given the difficulties in 
defining a reasonable variation in the theoretical framework, they are not
included in the uncertainties quoted on the final value of $\asmz$.

\section{Parton densities from ZEUS data alone}
\label{sec:special}

The fit using ZEUS data only (ZEUS-O) uses
the charged current  $e^+p$ data~\cite{epj:c12:411} and
the neutral and charged current $e^-p$ 
data~\cite{desy:02:113,pl:b539:197} in addition to the 
$e^+p$ neutral current data~\cite{epj:c21:443}.
These high-$Q^2$ data are very well described by the ZEUS-S fit, as 
illustrated in Figs.~\ref{fig:ncelpo} and~\ref{fig:ccelpo}. However,
in the ZEUS-O fit these additional data sets 
were used instead of the fixed-target 
data to constrain the valence distributions. Note that the $\chi^2$ for these
additional data sets
must be calculated by adding statistical and systematic errors in
quadrature. The correlated point-to-point systematic 
uncertainties are small compared to the statistical uncertainties for these
data sets.
Since the exclusion of the fixed-target data leaves no constraint on the
flavour content of the sea, the value of $p_1$ for the $\Delta$ 
distribution was fixed to the value determined in the ZEUS-S
fit. The value $\asmz = 0.118$ was fixed; all other parameters 
were varied as usual. 

\begin{figure}[!p]
\includegraphics[scale=.65]{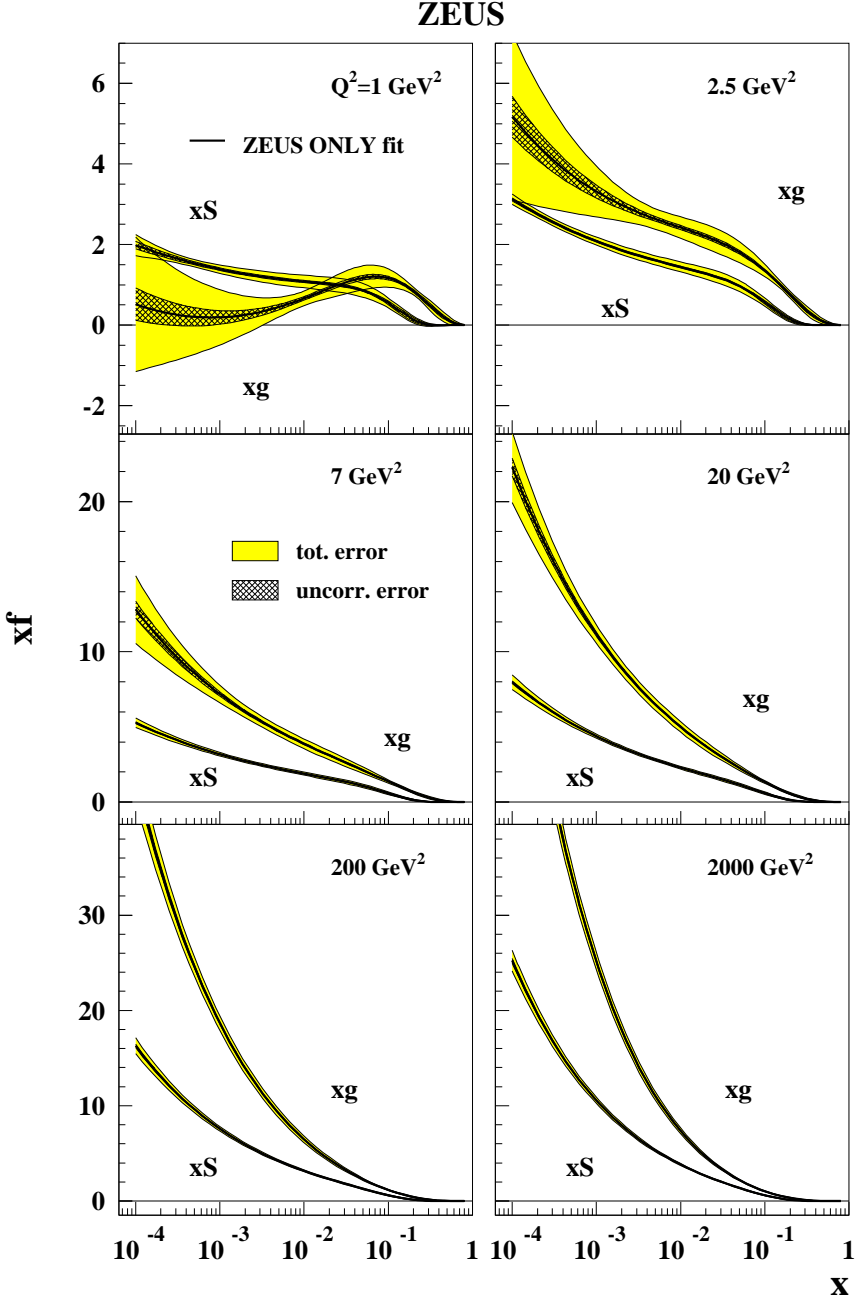}
\caption{The  gluon and sea distributions from the 
ZEUS-O NLO QCD fit in various
$Q^2$ bins. The error bands 
show the uncertainty from statistical and other uncorrelated sources
separately from the total uncertainty including correlated systematic 
uncertainties. The value of $\asmz = 0.118$ is fixed.
}
\label{fig:glusea_zo}
\end{figure}

The gluon and the sea distributions extracted from the 
ZEUS-O fit are shown in Fig.~\ref{fig:glusea_zo}. 
Comparing this figure to Fig.~\ref{fig:glusea}, it is clear that the
gluon and sea densities are mainly determined by the ZEUS data for 
$x < 10^{-2}$.
The ZEUS-O fit gives almost as good a determination of these 
distributions as the
ZEUS-S fit over most of the $x,Q^2$ plane used in the fit.

\begin{figure}[!p]
\includegraphics[scale=.75]{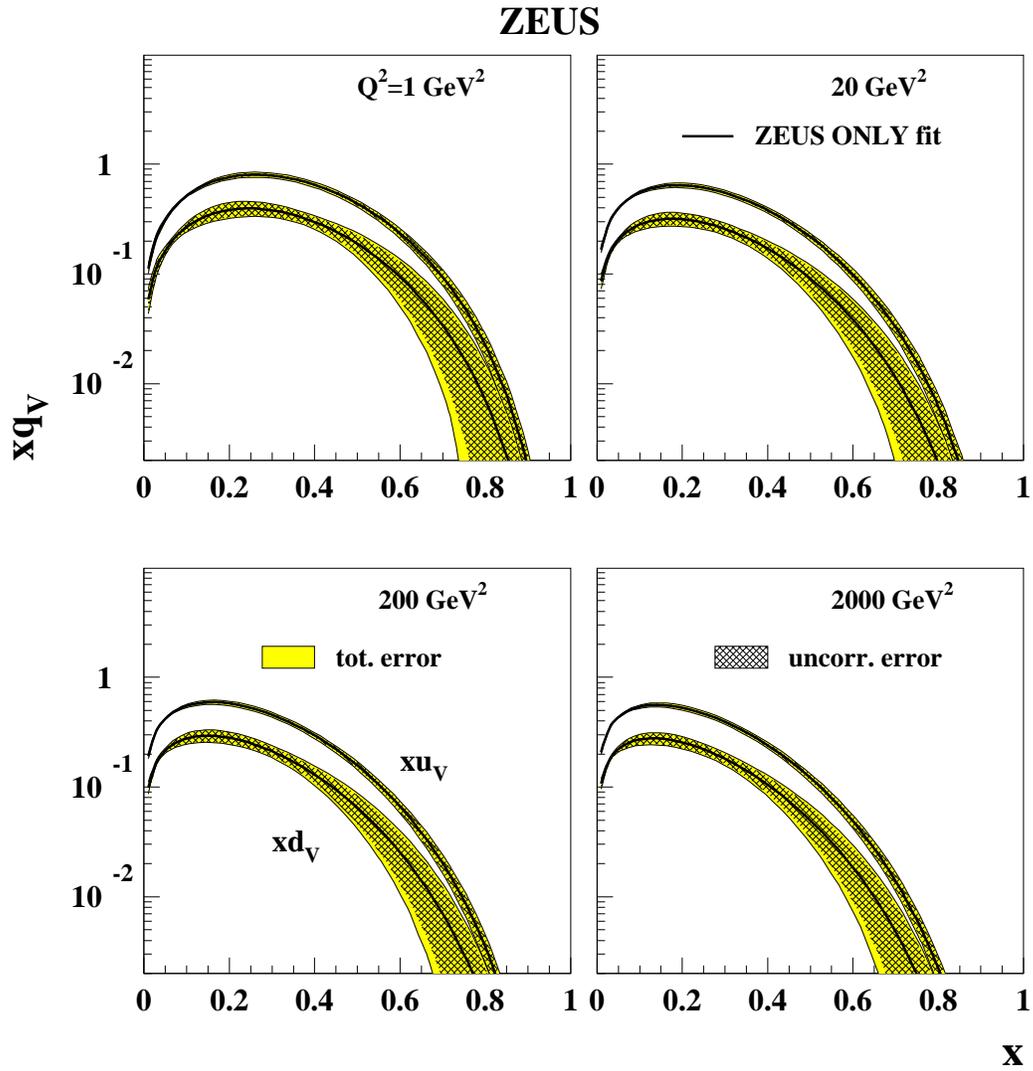}
\caption{The $xu_v$ and $xd_v$ distributions 
from the ZEUS-O NLO QCD fit.
The error bands are defined in the caption to Fig.~\ref{fig:glusea_zo}. 
The value of $\asmz = 0.118$ is fixed.
}
\label{fig:uvdv_zeus}
\end{figure}

The valence distributions extracted from the ZEUS-O fit are shown in 
Fig.~\ref{fig:uvdv_zeus}.
They are determined to a precision about a factor of two worse than in
the ZEUS-S fit. The $u$-valence distribution is well determined; 
 however, the $d$-valence distribution is much more poorly 
determined. In the ZEUS-O fit, the $d$-valence distribution is determined 
by the high-$Q^2$ $e^+p$ charged current data.
In contrast in the ZEUS-S fit the $d$-valence distribution is 
determined by the deuterium fixed-target data.
Recently it has been suggested that such measurements are 
subject to significant uncertainty from deuteron binding
corrections~\cite{pl:b377:11,pl:b400:220,prl:82:2467,epj:c13:241,prl:84:5456}.
The ZEUS-O extraction does not suffer
this uncertainty. It produces a larger $d$-valence 
distribution at high-$x$ than the ZEUS-S fit, as can be seen by
comparison with Fig.~\ref{fig:dv}, but there is no disagreement
within the limited statistical precision of the current high-$Q^2$ data.

\section{The transition to very low $Q^2$}
\label{sec:loq2qcd}

The ZEUS-S and ZEUS-$\alpha_s$ fits and the NLO QCD fits of 
MRST~\cite{pl:b387:419,epj:c4:463,npps:79:105,epj:c23:73} and 
CTEQ~\cite{pr:d55:1280,epj:c12:375,hep-ph-0201195}
give good descriptions
of $F_2$ data down to $Q^2$ values of $1-2\unit{GeV}^2$.
For such fits to be valid, it is necessary to assume that the
formalism is valid
even for low $Q^2$~\cite{np:a641:461}, 
where $\alpha_s$ is large and perturbation theory
may break down, as well as
for very low $x$, where $\ln(1/x)$ resummation terms should become
important~\cite{pr:d64:074005,hep-ph-0102151,pl:b474:372,np:b599:383,np:b621:359}.  
High-density and non-perturbative effects
\cite{ijmp:a13:3385} are also neglected.
To investigate if there is a low-$Q^2$ limit to the applicability of
the NLO QCD DGLAP formalism,
the ZEUS-S fit was extrapolated into the $Q^2$ region covered by ZEUS 
shifted-vertex (SVX) data~\cite{epj:c7:609} and the precise ZEUS 
beam-pipe-tracker (BPT) data~\cite{pl:b487:53}.

\begin{figure}[!p]
\begin{center}
\includegraphics[scale=.65]{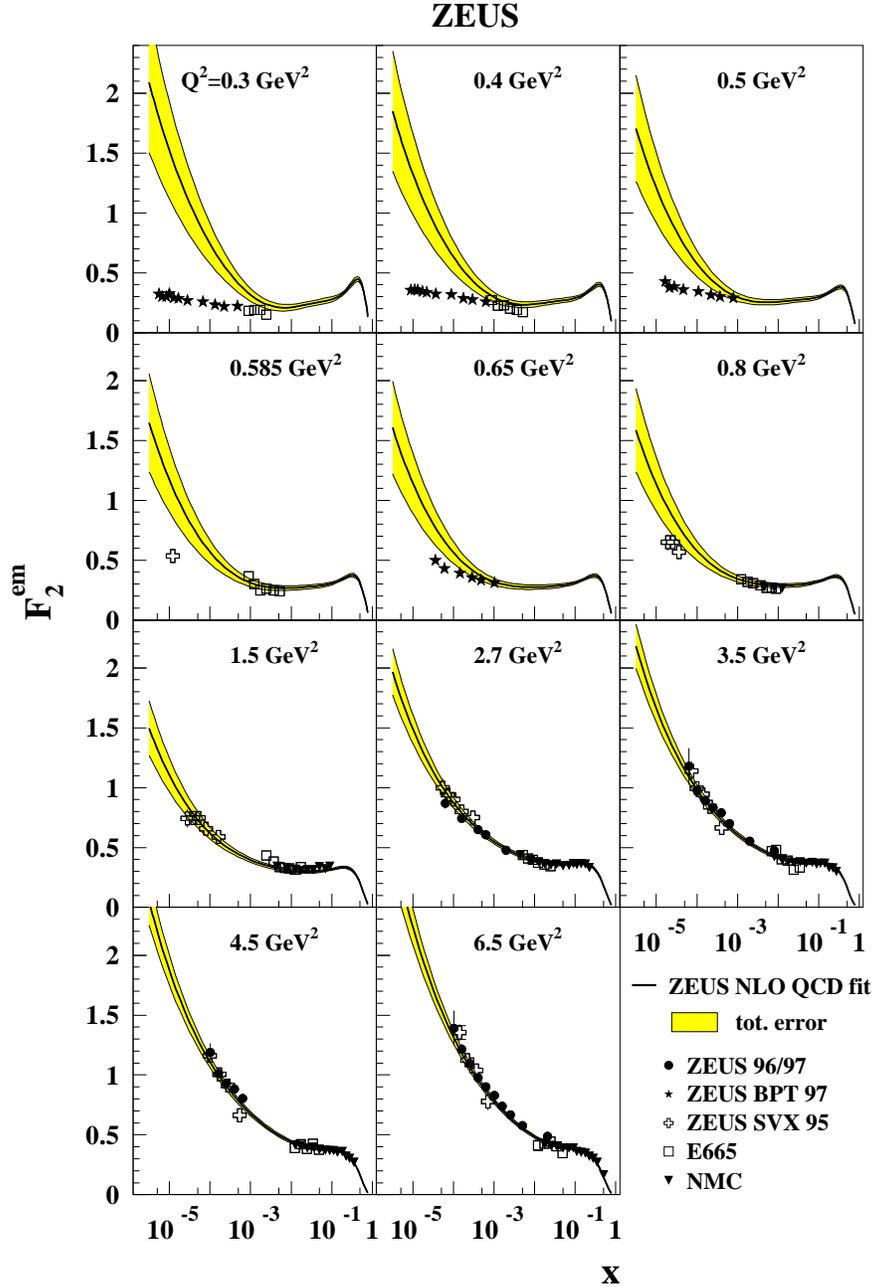}
\caption{$F_2$ data at very low $Q^2$ (including SVX95 and BPT97 data) compared
to the backward extrapolated ZEUS-S NLO QCD fit. 
The error bands are defined in the caption to Fig.~\ref{fig:f2lowall}.}
\label{fig:f2bpt}
\end{center}
\end{figure}

\begin{figure}[!p]
\begin{center}
\includegraphics[scale=.65]{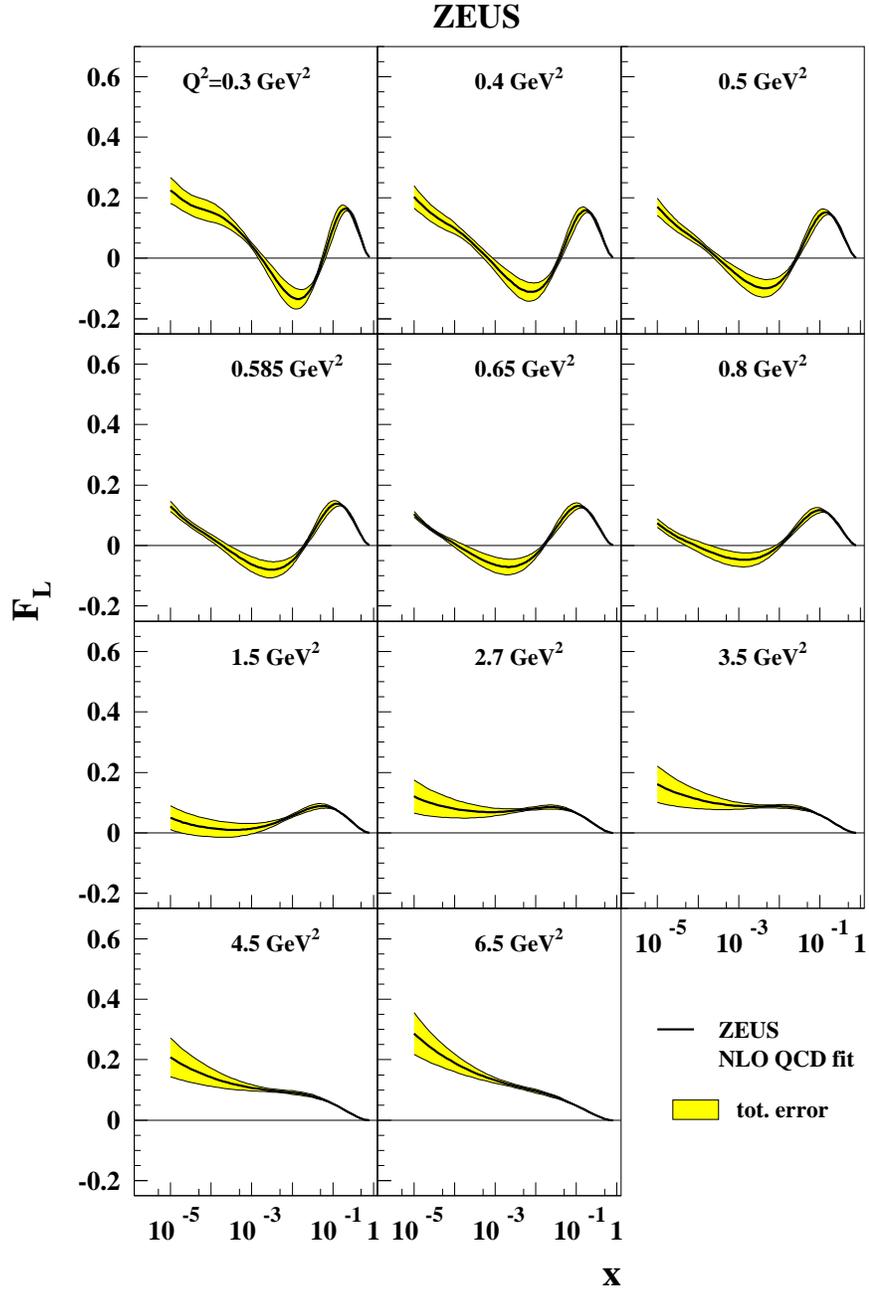}
\caption{The predictions for $F_L$ at very low $Q^2$ from the backward 
extrapolated ZEUS-S NLO QCD fit. The error bands are defined in the caption to
Fig.~\ref{fig:f2lowall}. 
}
\label{fig:flbpt}
\end{center}
\end{figure}

In Fig.~\ref{fig:f2bpt}, the ZEUS 96/97 data and the SVX and BPT data are 
shown at very low $Q^2$ compared to the predictions of the ZEUS-S fit. 
The increased precision of the new data, both at intermediate $Q^2$ and at 
very low $Q^2$, lead to a firmer conclusion than in the previous 
study~\cite{epj:c7:609}. The ZEUS-S fit is able to describe the data 
down to $Q^2 = 1.5~\unit{GeV}^2$, but exceeds the data at
$Q^2 = 0.8~\unit{GeV}^2$, and clearly fails for 
$Q^2~\leq~0.65~\unit{GeV}^2$, even when the conservative error bands 
on the fit due to the correlated systematic uncertainties are included. 
Thus, the NLO DGLAP formalism describes the extreme steepness of the 
ZEUS data at intermediate $Q^2$ ($2.7~\leq Q^2~\lesssim 200~\unit{GeV}^2$)
but is unable to accommodate the rapid transition to a flatter behaviour at 
$Q^2 < 1~\unit{GeV}^2$.
The ZEUS-S fit predictions for $F_L$ for very low-$Q^2$ values are also 
shown in Fig.~\ref{fig:flbpt}. The significantly negative values of
$F_L$ for $Q^2 \lesssim 1~\unit{GeV}^2$ are a further indication that
the NLO DGLAP formalism is not applicable. 

\section{Summary and conclusions}
\label{sec:concl}

The NLO DGLAP QCD formalism has been used to fit the 1996-1997 ZEUS data  
and fixed-target data in the kinematic region
$Q^2>2.5\unit{GeV}^2$, $6.3\times10^{-5}<x<0.65$ 
and $W^2 > 20\unit{GeV}^2$. Full account has been taken of correlated 
experimental systematic uncertainties.
A good description of the structure function and reduced
cross section over the $Q^2$ range from $2.5$ to
$30,000\unit{GeV}^2$ has been obtained. 

The parton distribution functions for the $u$ and $d$
valence quarks, the gluon and the total sea have been determined and the 
results are compatible with those of MRST2001 and CTEQ6. 
The ZEUS data are crucial
in determining the gluon and the sea distributions and a fit to ZEUS data
alone shows that these data also constrain the valence-quark
distributions. The new high-precision data
allow a greatly improved determination of the gluon and sea distributions. 
  
At $Q^2 \sim 1\unit{GeV}^2$, the fit predicts that the sea distribution 
is still rising at small $x$, whereas the gluon distribution is suppressed. 
The fit is unable to describe the precise ZEUS BPT data for
$Q^2 \lesssim 1~\unit{GeV}^2$ and 
also predicts unphysical negative values for 
$F_L$ in this $Q^2$ region. Hence the use of the NLO QCD DGLAP
formalism at $Q^2 \lesssim 1~\unit{GeV}^2$ is questionable.

The ZEUS data at low $x$ have been used to extract the 
value of $\asmz$ in a simultaneous fit
to $\asmz$ and the shapes of the input parton distributions, including
correlations between them, giving
\[
\asmz = 0.1166 \pm 0.0008(\rm uncorr.) \pm 0.0032(\rm corr.) \pm 0.0036(\rm norm.) \pm 0.0018(\rm model).
\]
 Uncertainties in the leading twist NLO QCD  
framework are also significant but cannot be easily quantified. 

The statistical accuracy of the ZEUS data is now sufficient to give a very good
determination of the sea and gluon PDFs. With the full HERA-II data sample, 
it will be possible to extend this analysis to give an accurate determination
of all proton PDFs within a single experiment.

\section*{Appendix}

\subsection*{A. Comparison of different ways of calculating $\chi^2$}

 The $\chi^2$ used in this analysis is defined in Eq.~(\ref{eq:chi2})
and the modification of the theoretical predictions to account for correlated
systematic uncertainties is given in Eq.~(\ref{eq:predmod}). 
The $\chi^2$ has been evaluated with the systematic-offset parameters set to
zero,
$s_\lambda=0$, with the consequence that the fitted theoretical predictions 
are as close as possible to the central values of the published data.
The offset parameters were then allowed to vary in the evaluation of the error
to account for correlation between systematic-uncertainty parameters 
and theoretical parameters, as described in Section~\ref{sec:errors}.

This method is referred to as the `offset method', since it is approximately 
equivalent to offsetting  each systematic parameter $s_\lambda$  by  
$\pm1$, performing a new fit for each of these variations
and adding in quadrature the resulting deviations of the theoretical 
parameters from their  central  values~\cite{jp:g28:779}. This
procedure does not assume that the systematic errors are
Gaussian distributed. This is a conservative method of error estimation
as compared to the Hessian methods described 
below~\cite{hep-ex-0005042,jp:g28:779}.

An alternative procedure would be to allow the systematic
uncertainty parameters $s_\lambda$ to vary in the main fit 
when determining the values of the
theoretical parameters. This was the procedure adopted by a recent H1 
analysis~\cite{epj:c21:33}, in which
only H1 and BCDMS data were considered. This method is referred to as `Hessian
method 1'.
The errors on the theoretical parameters are calculated from the
inverse of a single Hessian matrix which expresses the variation of 
$\chi^2$ with respect to both theoretical and systematic offset parameters.
Effectively, the theoretical prediction is not fitted
to the central values of the published experimental data, but allows 
these data points to move collectively according to their correlated
systematic uncertainties. The theoretical prediction determines the 
optimal settings for correlated systematic shifts of experimental data points 
such that the most consistent fit to all data sets is obtained. Thus
the fit correlates the systematic shifts in 
one experiment to those in another experiment.
 
Hessian method 1 becomes
a cumbersome procedure when the number of sources of systematic uncertainty
is large, as in  the present global DIS analysis. Recently the
CTEQ~\cite{pr:d65:014012, pr:d65:014013} collaboration has given an elegant 
analytical method for
performing the minimization with respect to systematic-uncertainty parameters.
This gives a new formulation of the $\chi^2$:
\[
\chi^2 = \sum_i \frac{\left[ F_i^{\rm NLOQCD}(p)-F_i(\rm meas) \right]^2}{(\sigma_{i,\rm stat}^2+\sigma_{i,\rm unc}^2)}  - B A^{-1} B,
\]
where
\[
B_\lambda = \sum_i \Delta_{i\lambda}^{\rm sys} \frac{\left[F_i^{\rm NLOQCD}(p)-F_i(\rm meas)\right]}{(\sigma_{i,\rm stat}^2+\sigma_{i,\rm unc}^2)}
\]
and
\[
A_{\lambda\nu} = \delta_{\lambda\nu} + \sum_i \Delta_{i\lambda}^{\rm sys} \Delta_{i\nu}^{\rm sys} /(\sigma_{i,\rm stat}^2+\sigma_{i,\rm unc}^2),
\]
such that the uncorrelated and correlated systematic 
contributions to the $\chi^2$ can be evaluated separately.
This method is referred to as `Hessian method 2'.

These two Hessian methods have been compared for the ZEUS-O fit,  
in which the systematic uncertainties are 
well understood. The results are very similar, 
as expected if the systematic uncertainties are Gaussian and the values
$\Delta^{sys}_{i\lambda}$ are standard deviations.
However, if data sets from different experiments are used in the fit, 
the results of these two
Hessian methods are only similar if normalisation uncertainties
are not included.

The offset method has been compared to Hessian method 2
by performing the ZEUS-$\alpha_s$ fit to global DIS data 
using Hessian method 2  to calculate the $\chi^2$.
Normalisation uncertainties were excluded and $\asmz$ was included 
as one of the theoretical parameters. 
This fit yields $\asmz = 0.1120 \pm 0.0013$,  where the error 
represents the total experimental uncertainty from correlated and uncorrelated
sources, excluding normalisation 
uncertainties.
Thus this value should be compared with $\asmz = 0.1166 \pm 0.0033$, 
evaluated  using the offset method, also excluding normalisation uncertainties
(see Eq.~\ref{eq:alphas}).  
Hessian method  2 gives a much reduced error estimate for both 
$\asmz$ and the PDF parameters.
The value of $\asmz$ is shifted from that obtained by the offset 
method. The PDF parameters are not affected
as strongly; their values are shifted by amounts which are 
well within the error 
estimates quoted for the offset method.

To compare the $\chi^2$ of the fits performed using the offset method and 
Hessian method 2, it is necessary to 
use a common method of $\chi^2$ calculation.
Table~\ref{tab:chibab} presents the $\chi^2$ for the theoretical parameters
 obtained using each of these methods, re-evaluated by adding 
statistical and systematic errors in quadrature.
For both methods, $\asmz$ has been included
among the theoretical parameters and normalisation uncertainties have not been
included among the systematic parameters. The total increase of $\chi^2$ for
Hessian method 2 as compared to the offset method is $\Delta \chi^2 = 283$.
Thus the results of Hessian method 2 represent a fit with an unacceptably large
value of $\chi^2$ 
when judged in this conventional way.

\subsection*{B. Parameter estimation and hypothesis testing}

To appreciate the signficance of the difference in $\chi^2$ between various
fits, the distinction between
the $\chi^2$ changes appropriate for parameter estimation and for 
hypothesis testing should be considered. 
Assuming that the experimental uncertainties that contribute have Gaussian
distributions, errors on theoretical 
parameters that are fitted within a fixed theoretical framework
are derived from the criterion for `parameter estimation' 
$\chi^2 \to \chi^2_{\rm min} + 1$. However, 
the goodness of fit of a theoretical
hypothesis is judged on the `hypothesis testing' criterion, such that
its $\chi^2$ should be approximately in the range $N \pm \sqrt (2N)$,
 where $N$ is the number of degrees of freedom. 

Fitting DIS data for PDF parameters and $\asmz$ is not a clean situation
of either parameter estimation or hypothesis testing, nor are the contributing
experimental uncertainties always Gaussian distributed.
Within the theoretical framework of leading-twist NLO QCD, 
many model inputs, such as the form of the PDF
 parameterisations, the values of cuts, the value of $Q^2_0$, the
data sets used in the fit, etc. can be varied.
These represent different hypotheses and they are accepted 
provided the fit $\chi^2$ falls within the hypothesis-testing criterion. The
theoretical 
parameters obtained for these different model hypotheses can differ from those
obtained in the standard fit by more than their errors as 
evaluated using the parameter-estimation criterion. In this case, the model
error on the parameters can exceed the estimate of the total experimental 
error. This does not happen for the
offset method, in which the uncorrelated experimental
errors evaluated by the parameter-estimation criterion are augmented by the 
contribution of the correlated experimental systematic
uncertainties, as explained in Section~\ref{sec:errors}. 
The shifts in 
theoretical parameter values for the different model hypotheses 
were found to be well within the total experimental error 
estimates\footnote{Note that 
this is true whether or not normalisation uncertainties
are included in these estimates.}. 
However, this is no longer the case when the
fit is performed using Hessian method 2. 

The CTEQ collaboration~\cite{pr:d65:014013,hep-ph-0201195} have considered 
this problem.
They consider that $\chi^2 \to \chi^2 + 1$ is not a reasonable tolerance
 on a global fit to approximately $1200$ data points from diverse sources, 
with 
theoretical and model uncertainties that are hard to quantify and experimental
uncertainties that may not be Gaussian distributed.
They have tried to formulate criteria for a more reasonable setting of the 
tolerance $T$, such that $\chi^2 \to \chi^2 + T^2$ becomes the variation 
on the basis of which errors on parameters are calculated. 
In setting this 
tolerance they have considered that all of the current 
world data sets must be acceptable and compatible at some level.
The level of tolerance they suggest is $ T \sim 10 $.
The error estimates of the present fit have been re-evaluated using
Hessian method 2 for various values of the tolerance. For $T=7$ the errors
on the PDF parameters and on $\asmz$ are very similar to those of the 
offset method performed under the same conditions~\footnote{This remains the
case when normalisation uncertainties are introduced into each of these 
methods. }. For example, 
the result $\asmz = 0.1120 \pm 0.0033$ is obtained.  Note that the value
$T = 7$  is similar to
the hypothesis-testing tolerance $T=\sqrt{\sqrt (2N)}$ for the fits.

Thus the offset method and the Hessian method with a modified
tolerance $T = 7$ give similar error
estimates. In choosing between these methods, there are some additional
considerations. In the Hessian method 2, it is necessary to check that data 
points are not shifted far outside their uncertainties. 
When the ZEUS-$\alpha_s$ and the ZEUS-S fit are done by Hessian method 2 
some of the systematic 
shifts for the ten classes of systematic uncertainty of the ZEUS data move by
$\sim \pm 1.4$ standard deviations. No single kinematic
region responsible for these shifts could be identified.
Whereas these shifts are not very large, 
it is significant that they differ from the systematic shifts to ZEUS data
determined in the CTEQ fit~\cite{hep-ph-0201195}. They also differ from 
those determined in the ZEUS-O fit done by Hessian method 2. 
Making different model
assumptions in the fits also produces somewhat different systematic shifts. 
It seems unreasonable to let variations in the model, or the choice of data 
included in the fit, change the best estimate
of the central value of the data points.

In summary, the offset method has been selected for several reasons.
First, its fit results make theoretical predictions that are 
as close
to the central values of the published data points as possible. 
The selection of data sets included in the fit or superficial changes
to the model do not change 
the best estimate of the central value of the data points. 
Secondly, it is approximately equivalent to a method that does not 
assume that experimental
systematic uncertainties are Gaussian distributed. Thirdly, its
results produce an acceptable
$\chi^2$ when re-evaluated conventionally by adding systematic and
statistical errors in quadrature.
Fourthly, its error estimates 
take account of the fact that the purpose is to estimate
errors on the PDF parameters and $\asmz$ within a general theoretical 
framework not specific to particular model choices. 
Quantitatively, the error 
estimates of the offset method correspond to those that would be obtained
using the more generous tolerance $T = 7$
in the more statistically powerful Hessian methods.

\section*{Acknowledgements}

We would like to thank R.S.~Thorne and R.G.~Roberts for useful discussions.
The experiment was made possible by the skill and dedication of the
HERA machine group who ran HERA most efficiently during the years
when the data used in this paper were collected. The realisation and
continuing operation of the ZEUS detector has been and is made possible by
the inventiveness and continuing hard work of many people not listed as
authors. Their contributions are acknowledged with great appreciation.
The support and encouragement of the DESY directorate continues to be
invaluable for the successful operation of the ZEUS collaboration. 

We acknowledge support by the following: the Natural Sciences and Engineering 
Research Council of Canada (NSERC); the German Federal Ministry for
Education and Research (BMBF), under contract numbers HZ1GUA 2, HZ1GUB 0, HZ1PDA 5, HZ1VFA 5;         
the MINERVA Gesellschaft f\"ur Forschung GmbH; the Israel Science Foundation, the U.S.-Israel Binational 
Science Foundation; the Israel Ministry of Science and the Benozyio Center for High Energy Physics;
the German-Israeli Foundation, the Israel Science Foundation; the Italian National Institute for Nuclear 
Physics (INFN); the Japanese Ministry of Education, Science and Culture (the Monbusho) and its grants for 
Scientific Research; the Korean Ministry of Education and Korea Science and Engineering Foundation;            
the Netherlands Foundation for Research on Matter (FOM); the Polish State Committee for Scientific
Research, under grant numbers 620/E-77/SPUB-M/DESY/P-03/DZ 247/2000-2002; the Fund for Fundamental
Research of Russian Ministry for Science and Education; the Spanish Ministry of Education and Science
through funds provided by CICYT; the Particle Physics and Astronomy Research Council, UK; the US
Department of Energy; the US National Science Foundation; the Polish State Committee for Scientific
Research under grant numbers 112/E-356/SPUB-M/DESY/P-03/DZ 301/2000-2002, 2 P03B 13922, 
115/E-343/SPUB-M/DESY/P-03/DZ 121/2001-2002, 2 P03B 07022.                    
J. Rautenberg was supported by the German-Israeli Foundation, contract
number I-523-13.7/97; D.S.~Bailey and N.H.~Brook are PPARC Advanced
fellows; E.~Rodrigues and R.~Gon\c{c}alo were supported by the Portuguese Foundation for
Science and Technology; J.~Szuba was partly supported by the Israel Science Foundation and
the Israel Ministry of Science; A.~Kota\'{n}ski was supported by the Polish State
Committee for Scientific Research, grant number 2 P03B 09322; J.H.~Loizides was supported
by Argonne National Laboratory, USA; R.~Ciesielski was supported by the Polish State
Committee for Scientific Research, grant number 2 P03B 07222; T.~Tymieniecka and A.~Ukleja were
supported by German Federal Ministry for Education and Research (BMBF), POL 01/043;
L.K.~Gladilin was partly supported by University of Wisconsin via the U.S.-Israel Binational
Science Foundation.

{
\def\bibname{\Large\bf References}
\def\refname{\Large\bf References}
\pagestyle{plain}
\ifzeusbst
  \bibliographystyle{../BiBTeX/bst/l4z_default}
\fi
\ifzdrftbst
  \bibliographystyle{../BiBTeX/bst/l4z_draft}
\fi
\ifzbstepj
  \bibliographystyle{../BiBTeX/bst/l4z_epj}
\fi
\ifzbstnp
  \bibliographystyle{../BiBTeX/bst/l4z_np}
\fi
\ifzbstpl
  \bibliographystyle{../BiBTeX/bst/l4z_pl}
\fi
{\raggedright
\bibliography{../BiBTeX/bib/l4z_articles,%
              ../BiBTeX/bib/l4z_books,%
              ../BiBTeX/bib/l4z_conferences,%
              ../BiBTeX/bib/l4z_h1,%
              ../BiBTeX/bib/l4z_misc,%
              ../BiBTeX/bib/l4z_old,%
              ../BiBTeX/bib/l4z_preprints,%
              ../BiBTeX/bib/l4z_replaced,%
              ../BiBTeX/bib/l4z_temporary,%
              ../BiBTeX/bib/l4z_zeus}}
}
\vfill\eject

%
%
\newpage
\begin{table}
\begin{center}
\begin{tabular}{|l|c|c|c|c|}
\hline
\footnotesize{Experiment} & \footnotesize{Number of data} & \footnotesize{$\chi^2$ per} & \footnotesize{Number of correlated}  & \footnotesize{Number of normalisation}   \\
           & \footnotesize{points}  & \footnotesize{data point}        & \footnotesize{systematic uncertainties} & \footnotesize{uncertainties} \\
\hline
\footnotesize{ZEUS96/97~\cite{epj:c21:443}} & 242   & 0.85 & 10 & 2\\
\hline
\footnotesize{BCDMS p~\cite{pl:b223:485}}   & 305   & 0.94 &  5 & 5\\
\hline
\footnotesize{NMC p~\cite{np:b483:3}}       & 218   & 1.21 &  \lw{12} & \lw{4}\\
\footnotesize{NMC D~\cite{np:b483:3}}       & 218   & 0.92 &    &  \\
\hline
\footnotesize{NMC D/p~\cite{np:b487:3}}     & 129   & 0.94 &  5 & 0\\
\hline
\footnotesize{E665 D~\cite{pr:d54:3006}}    &  47   & 0.94 & \lw{7} & 2\\
\footnotesize{E665 p~\cite{pr:d54:3006}}    &  47   & 1.16 &    & 1\\
\hline
\footnotesize{CCFRxF3~\cite{prl:79:1213}}   &  57   & 0.40 & 18 & 0\\
\hline
\end{tabular}
\caption{Table of $\chi^2$ for the data sets used in the ZEUS-S 
NLO QCD fit, evaluated by adding all systematic and statistical 
uncertainties in quadrature. Note that for CCFR data no separate total 
systematic uncertainty is supplied, so that this procedure overestimates 
the total uncertainty. The number of correlated 
systematic uncertainties for each data set is also given. 
Note that the systematic uncertainties for the p and D data sets of NMC 
and E665 must be taken together.
The normalisations of the four beam energies comprising the 
NMC data are the same for the p and D targets, whereas for E665 data 
there is a separate normalisation uncertainty for the p and D targets as well
as a common normalisation uncertainty. The number of normalisation 
uncertainties for BCDMS data derives from the four beam energies of 
the data and an overall normalisation uncertainty. There are two ZEUS 
normalisation
uncertainties: an overall uncertainty and the relative uncertainty of the
data for which $Q^2 < 30\unit{GeV}^2$, with respect to the higher $Q^2$ data. 
The CCFR normalisation uncertainty is included amongst its
systematic uncertainties.
   }

\label{tab:tabchi}
\end{center}
\end{table}

\clearpage
\begin{table}
\begin{center}
\begin{tabular}{|l|c|c|c|c|}
\hline
PDF & $p_1$& $p_2$ & $p_3$ & $p_5$ \\
\hline
 $xu_v$ &    \footnotesize{($1.69 \pm 0.01 \pm 0.06$)} &
             \footnotesize{$0.5$}                      &
             \footnotesize{$4.00 \pm 0.01 \pm 0.08$}   &
             \footnotesize{$5.04 \pm 0.09 \pm 0.64$}   \\
\hline
 $xd_v$ &
             \footnotesize{($0.96 \pm 0.01  \pm 0.08$} &
             \footnotesize{$0.5$}                      &
             \footnotesize{$5.33 \pm 0.09 \pm 0.48$}   &
             \footnotesize{$6.2 \pm 0.4 \pm 2.3$}      \\
\hline
 $xS$ &      \footnotesize{$0.603 \pm 0.007 \pm 0.048$} &
             \footnotesize{$-0.235 \pm 0.002 \pm 0.012$}&
             \footnotesize{$ 8.9 \pm 0.2 \pm 1.2$}     &
             \footnotesize{$6.8 \pm 0.4 \pm 2.0$}      \\
\hline
 $xg$  &     \footnotesize{($1.77 \pm 0.09 \pm 0.49$)} &
             \footnotesize{$-0.20 \pm 0.01 \pm 0.04$}  &
             \footnotesize{$6.2 \pm 0.2 \pm 1.2$}      &
             \footnotesize{$0$}                        \\
\hline
 $x\Delta$ & \footnotesize{$0.27 \pm 0.01 \pm 0.06$}  &
             \footnotesize{$0.5$}                     &
             \footnotesize{($10.9 \pm 0.2 \pm 1.2$)}  &
             \footnotesize{$0$}                       \\
\hline
\end{tabular}
\caption{Table of PDF parameters at $Q^2_0$, as determined from
the ZEUS-S fit. The first uncertainty given derives from statistical
and other uncorrelated sources and the second uncertainty is the
additional
contribution from correlated systematic uncertainties.
The numbers in parentheses were derived from the fitted parameters as
described
in the text.
}
\label{tab:param}
\end{center}
\end{table}

\clearpage
\begin{table}
\begin{center}
\begin{tabular}{|l|c|c|}
\hline
Nominal value & New value& $\Delta \asmz$  \\

\hline
$Q^2_0= 7\unit{GeV}^2$& $Q^2_0= 4\unit{GeV}^2$ & +0.0008 \\
        & $Q^2_0=10\unit{GeV}^2$ & -0.0004\\
\hline
$Q^2_{\rm min}=2.5\unit{GeV}^2$  
     &  $Q^2_{\rm min}=4.5\unit{GeV}^2$ & -0.0007\\
\hline
 $x_{\rm min}=6.3\times 10^{-5}$ 
                         &  $x_{\rm min}=10^{-3}$ & -0.0005\\
\hline
 $W^2_{\rm min}=20\unit{GeV}^2$  &  $W^2_{\rm min}=10\unit{GeV}^2 $ & -0.0005\\

\hline
\end{tabular}
\caption{Shifts in the central value of $\asmz$ with 
variation of the nominal input values for the ZEUS-$\alpha_s$ fit.} 
\label{tab:model}
\end{center}
\end{table}

\begin{table}
\begin{center}
\begin{tabular}{|l|c|c|c|}
\hline
Experiment & Data points& $\chi^2$/data point& $\chi^2$/data point \\
           &            & Hessian method 2 & Offset method \\

\hline
 ZEUS96/97 & 242 & 1.37 & 0.83\\
 BCDMS p   & 305 & 0.95 & 0.89 \\ 
 NMC p     & 218 & 1.50 & 1.26\\
 NMC D     & 218 & 1.15 & 0.96\\
 NMC D/p    & 129   & 0.97& 0.93 \\
 E665 D    &  47   & 0.97 & 0.94 \\
 E665 p    &  47   & 1.17 & 1.16 \\
 CCFRxF3   &  57   & 0.99 & 0.39 \\

\hline
\end{tabular}
\caption{Table of $\chi^2$ calculated by adding systematic and statistical 
errors in quadrature for the theoretical parameters determined by the 
offset method and Hessian method 2}
\label{tab:chibab}
\end{center}
\end{table}

\clearpage

\clearpage

\clearpage

\clearpage

\clearpage

\clearpage

\clearpage

\clearpage

\clearpage

\clearpage

\clearpage

\clearpage

\clearpage

\clearpage

\clearpage

\clearpage


\end{document}